\documentclass[dvipsnames,tikz]{acmart}
\usepackage{pgfplots}
\usepackage{pgf-pie}
\usetikzlibrary{patterns}
\usetikzlibrary{backgrounds}
\usetikzlibrary{shapes}
\pgfplotsset{compat=1.16}
\usepackage{censor}
\usepackage{subfig}
\AtBeginDocument{%
  }

\tikzset{
  chart/.style={
    legend label/.style={font={\scriptsize},anchor=west,align=left},
    legend box/.style={rectangle, draw, minimum size=5pt},
    axis/.style={black,semithick,->},
    axis label/.style={anchor=east,font={\tiny}},
  },
  bar chart/.style={
    chart,
    bar width/.code={
        \pgfmathparse{##1/2}
        \global\let\bar@w\pgfmathresult
    },
    bar/.style={very thick, draw=white},
    bar label/.style={font={\bfseries\small},anchor=north},
    bar value/.style={font={\footnotesize}},
    bar width=.75,
  },
  pie chart/.style={
    chart,
    slice/.style={line cap=round, line join=round, thin, draw=black},
    pie title/.style={},
    slice type/.style n args={3}{
        ##1/.style={pattern color=##2,pattern=##3},
        values of ##1/.style={}
    }
}}
\pgfdeclarelayer{background}
\pgfdeclarelayer{foreground}
\pgfsetlayers{background,main,foreground}

\newcommand{\pie}[3][]{
    \begin{scope}[#1]
    \pgfmathsetmacro{\curA}{90}
    \pgfmathsetmacro{\r}{1}
    \def\c{(0,0)}
    \node[pie title] at (90:1.3) {#2};
    \foreach \v/\s in{#3}{
        \pgfmathsetmacro{\deltaA}{\v/100*360}
        \pgfmathsetmacro{\nextA}{\curA + \deltaA}
        \pgfmathsetmacro{\midA}{(\curA+\nextA)/2}

        \path[slice,\s] \c
            -- +(\curA:\r)
            arc (\curA:\nextA:\r)
            -- cycle;
        \pgfmathsetmacro{\d}{max((\deltaA * -(.5/50) + 1) , .5)}

        \begin{pgfonlayer}{foreground}
        \path \c -- node[pos=\d,pie values,values of \s]{$\v\%$} +(\midA:\r);
        \end{pgfonlayer}

        \global\let\curA\nextA
    }
    \end{scope}
}

\begin{document}

\title{Scale-Score: Investigation of a Meta yet Multi-level Label to Support Nutritious \\and Sustainable Food Choices When Online Grocery Shopping}

\author{Marco Druschba}
\email{marco.druschba@uni-oldenburg.de}
% \orcid{0000-0002-7634-9482}
\affiliation{%
  \institution{University of Oldenburg}
  %\streetaddress{P.O. Box 1212}
  \city{Oldenburg}
  \country{Germany}
}

\author{G\"ozel Shakeri}
% \orcid{0000-0002-3154-0814}
\affiliation{%
  \institution{University of Oldenburg}
  \city{Oldenburg}
  \country{Germany}}
\email{gozel.shakeri@uni-oldenburg.de}

% \author{Susanne Boll}
% % \orcid{0000-0003-4293-1623}
% \affiliation{%
%  \institution{University of Oldenburg}
%  \city{Oldenburg}
%  %\state{Arunachal Pradesh}
%  \country{Germany}}

% \author{Anonymous Author}
% % \orcid{0000-0003-4293-1623}
% \affiliation{%
%  \institution{University of Earth}
%  \city{City}
%  %\state{Arunachal Pradesh}
%  \country{Earth}}

\renewcommand{\shortauthors}{Druschba et al.}

\begin{abstract}
  Food consumption is one of the biggest contributors to climate change. However, online grocery shoppers often lack the time, motivation, or knowledge to contemplate a food's environmental impact. At the same time, they are concerned with their own well-being. To empower grocery shoppers in making nutritionally and environmentally informed decisions, we investigate the efficacy of the Scale-Score, a label combining nutritional and environmental information to highlight a product's benefit to both the consumer's and the planet's health, without obscuring either information. We conducted an online survey to understand user needs and requirements regarding a joint food label, we developed an open-source mock online grocery environment, and assessed label efficacy. We find that the Scale-Score supports nutritious purchases, yet needs improving regarding sustainability support. Our research shows first insights into design considerations and performance of a combined yet disjoint food label, potentially altering the label design space.
\end{abstract}

\begin{teaserfigure}
    \centering
    \includegraphics[width=\textwidth]{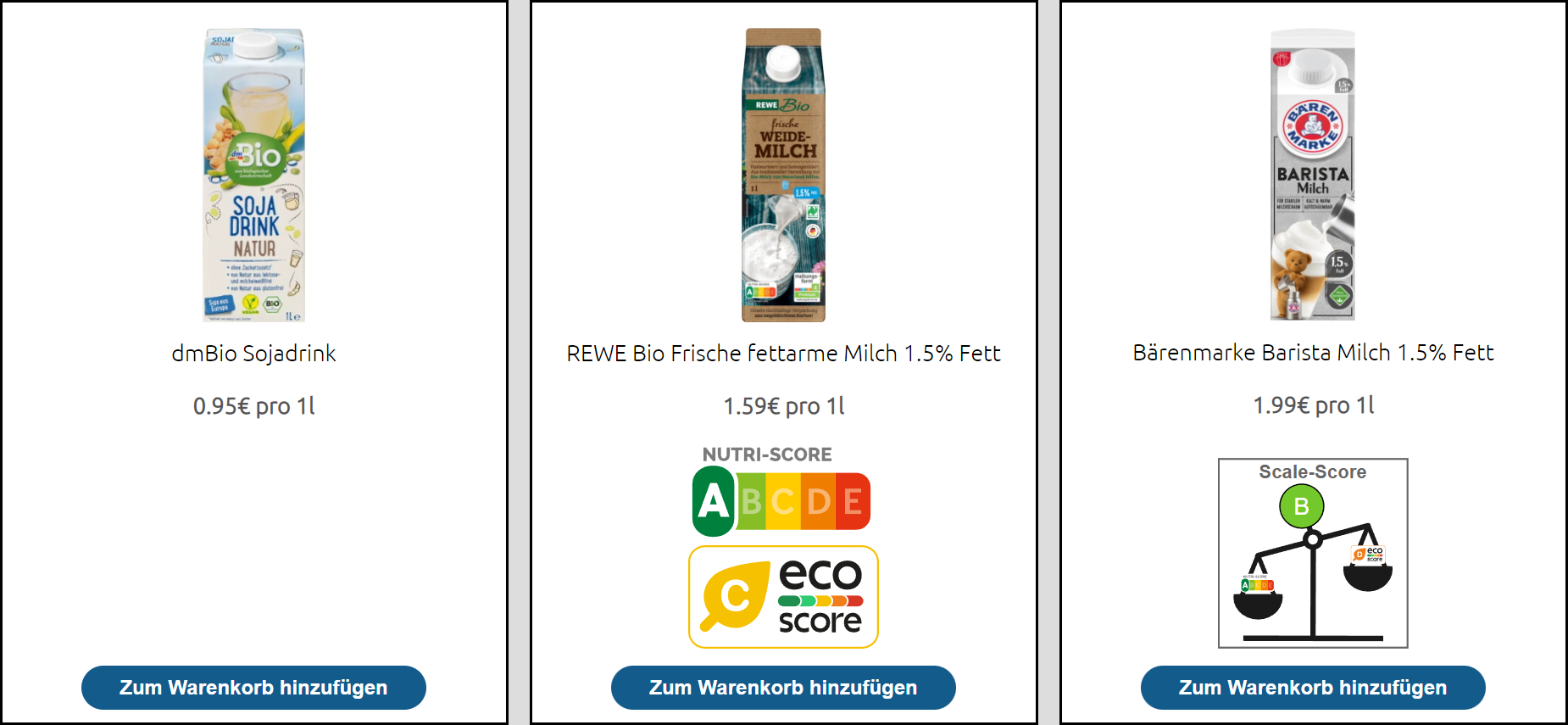}
    \caption{Scale-Score (right) provides people with high level information about the nutritional and sustainable value of foods (i.e. meta-label), yet gives additional information to allow for individual prioritisation of health or environment (multi-level label). In a mock-up online supermarket we tested three conditions (left: no labels, middle: Nutri- and Eco-Score, right: Scale-Score) showing that the Scale-Score supports nutritious purchase decisions, yet needs improving regarding sustainability support.}
    \label{fig:all}
    \Description{Figure 1 shows an excerpt of a screenshot of a developed online store. Next to each other, there are three milk products visible, each of them marked with different food labels. Product 1 (left) does not have an additional label. Product 2 (middle) is labelled with its corresponding Nutri-Score and Eco-Score. Product 3 (right) is marked with a designed Scale-Score, a multi-level label for additional health and environmental information. For each product a picture, it's brand, product name and price is also given.}
\end{teaserfigure}

\maketitle

\section{Introduction}
%
% Questions for good introduction:
%
% 1) What is the larger scope and problem space? And why should we care?
% 2) What is the specific problem addressed?
% 3) Why is the problem important? Why was this work carried out?
% 4) What have you done?
% 5) What is new about your work?
% 6) What did you find out? What are the concrete results?
% 7) What are the implications on a larger scale? How does it change the bigger picture? 
% 

The food sector is a major contributor to climate change, responsible for one-third of global greenhouse gas emissions \cite{Poore2018ReducingFoodImpact}. %To avoid the worst effects of a changing climate, it is necessary to significantly reduce the food sector's environmental impact. However, the food sector's emissions are expected to increase \cite{Msangi2012_FeedingFuturesChangingDiets}, rising from 30\% of global greenhouse gas emissions in 2020 to 50\% by 2050 \cite{Springmann2016_AnalysisValuationHealthClimateChangeCobenefitsDietaryChange}. 
Research has shown that shifting to a diet that includes low-impact foods like plants can reduce the food sector's emissions by half \cite{scarborough2014dietary}. %This could be a quick way to significantly reduce global greenhouse gas emissions and buy time for more sustainable policies \cite{Thomas2017_HCIEnvironmentalPublicPolicy} and changes to production and supply chains \cite{Poore2018ReducingFoodImpact}. 
However, there are many barriers that prevent consumers from making this shift, including a lack of knowledge about the environmental impact of individual food choices \cite{MACDIARMID2016EatingLikeTheresNoTomorrow, Nguyen2019_GreenConsumptionClosingIntentionBehaviorGap,Hanss2020_PerceivedConsumerEffectiveness} and which foods are considered sustainable \cite{Poore2018ReducingFoodImpact, LENTZ2018GaugingAttitudesMeatConsumption}. This is known as the intention-behaviour gap \cite{Vermeir2006_SustainableFoodConsumptionExploringConsumerAttitudeBehavioralIntentionGap} which exists for people who want to consume more sustainably but cannot translate their intentions into actions.

Labels help overcome the intention-behaviour gap by transforming complex information about food e.g., nutritional values, animal welfare standards, or environmental aspects, into simple logos or diagrams \cite{asioli2020sustainability, Dufeu2014multilabellingandWTP, lemken2021climatescorelabel, kaczorowska2019perceivedproductvalue}. Food-label technology and personal informatics thereby both use similar techniques to motivate users \cite{sauve2020econundrum}, such as providing information, enabling comparison, and giving feedback \cite{Froehlich2010ecotechnology}. These technologies can use either abstract or concrete information, presented visually or through text, to educate users and encourage sustainable or nutritious shopping behaviour. Several studies within the HCI discipline investigated labels as a means of providing education tailored to users' own context and choices \cite{Froehlich2010ecotechnology, Starke2017_EffectiveUserInterfaceDesignsIncreaseEnergyEfficientBehaviour, SANGUINETTI2018_DesignBehaviourFrameworkEffectivenessEcoFeedback}. HCI examples of eco-labels on food are EcoPanel \cite{zapico2016ecopanel}, Social Recipes \cite{yalvac2014socialrecipes}, Envirofy \cite{shakeri2021envirofy}, Nu-Food \cite{PANZONE2021nufood}, Econundrum \cite{sauve2020econundrum}, and Food Qualculator \cite{clear2012designing}. 

Developing solutions which support people in making decisions which are `good for the environment' must equate to being `good for the body'. A plethora of research has shown that our health is impacted by the foods we consume, and following a balanced sustainable diet --- i.e. a plant-based diet (low frequency consumption of animal products) may be key to better bodily and planetary health. It improves physiological health \cite{dinu2017healthyvegan}, reduces risks of nutrition-related diseases (e.g. coronary heart disease \cite{Satija2017veganheartdisease}), and lowers environmental impacts  significantly \cite{Poore2018ReducingFoodImpact}. HCI examples of effective nutrition labels are BetterChoice \cite{fuchs2022betterchoice}, FLICC \cite{Harrington2019Flicc}, and HealthierU \cite{Luo2017_OnlineEngagementWebbaseSupermarketHealthProgram}. % Yet, supporting consumers in both, sustainable and healthy food choices is surprisingly under-explored in HCI.

Although health- and environmental challenges are often addressed separately, they are closely intertwined and surprisingly under-explored in HCI. There is an urgent need to promote sustainable diets more effectively, when it matters most, \textit{while} shopping \cite{zapico2016ecopanel, Terzimehi2017_ChallengesJustInTimeAdaptiveFoodChoiceInterventions, Luo2017_OnlineEngagementWebbaseSupermarketHealthProgram}. A major challenge in creating a solution (e.g. label) which communicates health \textit{and} environmental concerns is, however, that visualisations of nutritious and sustainable foods are lived differently. Communication of nutritious foods uses \textit{meta} labels \cite{dendler2014metalabelling, TORMA2021metalabelling}, which aim to combine multiple dimensions into a single label to reduce the number of labels on a product and help consumers make more sustainable choices (e.g. Nutri-Score \cite{nslogo}). Signalling sustainability however, relies on \textit{multi-level} labels which only address a single dimension such as animal welfare labels, organic certification, and environmental labels, etc. \cite{ISABELSONNTAG2023labeljungle}. These are more popular as well as more prevalent on the sustainability label market, with 58 different food eco-labels (as of January 2023) \cite{EcolabelIndex}.

While multi-level labels allow for preference-based choices, they may lead to information overload \cite{yokessa:hal-02628579} requiring consumers to choose between different labels that address different sustainability dimensions (i.e. intra-sustainability trade-offs \cite{luchs2017tradeoffs}). This in turn can result in consumers either ignoring any further information and basing their decision on price only \cite{APOSTOLIDIS2019109, GrebitusRoosenSeitz+2015+73+81, Simeone2016labelcomplexity}, decreasing consumer trust and label credibility \cite{kaczorowska2019perceivedproductvalue, MILLER2015207}, or in over-evaluating positive attributes of the labels \cite{APOSTOLIDIS2019109}. Displaying two or more labels and thereby triggering intra-sustainability trade-offs \cite{luchs2017tradeoffs} is a strategy which confuses, overwhelms and ultimately fails well-intended consumers. A balance must be found between 1) a \textit{meta-label} which summarises the nutritional and environmental information into one piece of information, digestible at a single glance and 2) a \textit{multi-level label} which allows for preference-based decision making. Reflecting on the limitations of current label research, providing a single meta yet multi-level label may be key to healthy and eco-friendly purchases. 

Our research focuses on the design space of labels which comprise of both, health and environmental information; when food shopping online. Online grocery shopping is growing rapidly \cite{InternetStatistics}. While online shopping is convenient, accessible, fast, and generally more sustainable than in-person shopping due to lower transportation emissions \cite{CARLING2015onlinetransportCO2}, there is still room to improve it. The advantage of conducting research in online environments is that supermarket websites can be easily and flexibly modified (e.g. through web browser extensions \cite{shakeri2021envirofy}). This approach allows for the early and frequent sharing of results, bridging the gap between theory and practice quickly. This is important in order to support agile and responsive research \cite{Hekler2016_AgileScienceCreatingUsefulProducts} that stays current with and takes advantage of the latest technology \cite{Hekler2013_MindTheTheoreticalGap}. Online grocery shopping presents an opportunity for sustainable HCI to transform food label research (i.e. data visualisation, personal informatics) towards faster, more agile, and technology-mediated research practice. 

This paper outlines the multidisciplinary development of a combined yet disjoint ecological and nutritious label i.e. the Scale-Score (Figure \ref{fig:all}, right). We describe 1) the user-centred design of a combining label, i.e. the Scale-Score, 2) the design of the mock online grocery store, and 3) a within-subjects study with a representative sample of 12 German consumers making food choices in said mock online store. The experiment tested the impact of presenting the Scale-Score on the nutritional quality and environmental impact of the consumers' food choices, compared to the effects of both Nutri-Score and Eco-Score labels, and no persuasive technology. %The study calculated the average nutritional quality and environmental impact of the food choices made by the participants in order to evaluate the effects of the various labelling conditions. 
The results show that displaying both Nutri-Score and Eco-Score labels improved the nutritional as well as environmental values per basket. The Scale-Score improved nutritional quality of purchases, however surprisingly, it performed worse in terms of environmental impact, compared to Nutri-Eco and baseline condition. The poorer environmental performance of the Scale-Score was largely due to its design: nutritional scores weigh in more towards the final score compared to the environmental data. Overall, this paper makes two contributions: 1) it provides first evidence in support of using a joint yet disjoint nutritional and ecological label to encourage transitions towards healthier and more sustainable diets, when online shopping; 2) it makes the mock online grocery store --- an environment developed for research purposes --- available for other researchers to use and modify freely, encouraging fast and agile research practices. 

\section{Background}
% 
% Points to (potentially) include in Related Work
% 
% 1) Work that proposes a different method to solve the same problem.
% 2) Work that uses the same proposed method to solve a different problem.
% 3) A method that is similar to your method that solves a relatively similar problem.
% 4) A discussion of a set of related problems that covers your problem domain.
%

To empower grocery shoppers in making the `right' choice, it is important to communicate the nutritional and environmental information in an effective way. In and outside of HCI, the most common approach are food labels. The basic idea of a (front-of-package) label is to present complex information about a product's characteristics, such as nutritional values, animal welfare standards, and environmental impacts, in a simplified form to make it easier for consumers to make informed decisions \cite{lemken2021climatescorelabel, kaczorowska2019perceivedproductvalue, Dufeu2014multilabellingandWTP,asioli2020sustainability}. 

\subsection{Meta-Labels}
\label{subsec:meta-labels}

In North-West Europe, the two most prevalent nutrition and environment food meta-labels are the Nutri-Score \cite{nslogo} and the Eco-Score \cite{eslogo}, respectively. Both are front-of-package labels using the same visualisation vocabulary; colour-coding system to provide consumers with a uniform, easy-to-understand summary of a food's nutritional value or ecological impact; five-colour scale, with corresponding letters, in order to make it accessible and easy to understand. 

\begin{figure}[h!]%
    \centering
    \subfloat[\centering Nutri-Score \cite{nslogo} label\label{fig:nutri-ex}]{{\includegraphics[width=3cm]{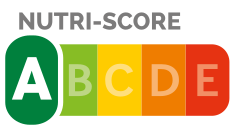}}}%
    \qquad
    \subfloat[\centering Eco-Score \cite{eslogo} label\label{fig:eco-ex}]{{\includegraphics[width=3cm]{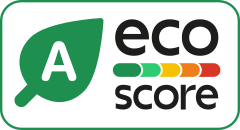}}}% 
    \caption{Nutrition and environment food meta-labels.}%
    \label{fig:nutri-eco}%
    \Description{Figure 2 labelled `(a)' shows an exemplary Nutri-Score label with rating `A'. Figure 2 labelled `(b)' shows an exemplary Eco-Score label with rating `A'.}
\end{figure}

Neither the Nutri-Score (Figure \ref{fig:nutri-ex}) nor the Eco-Score categories foods as `healthy' or `unhealthy', `sustainable' or `unsustainable', but rather provide semi-quantitative information \cite{hercberg2021nutri} about the relative nutritional or environmental value of a product compared to similar products \textit{within} a food category (e.g. pizza). This helps consumers make better choices within the category. Over the years, a large body of literature has demonstrated the effectiveness of the Nutri-Score label, particularly when compared to other labels, in terms of its ability to inform and influence consumer behaviour \cite{BMLnutriscore,HISPACOOP,TestAchats,DETEMMERMAN2021104995,Barthelemy2020nutriscoreFrance, egnell2018objective}. Meta-Labels are persuasive technologies guiding people to conform with normative consumerism. 

\subsection{Multi-level Labels}

Multi-level labels allow for critical consumerism. The availability of diverse labels gives people the freedom to make choices on various issues of the food system (e.g. health, environment) \cite{lawo2021criticalconsumer}, supporting the personalisation of what are the `right' values for consumers, rather than enforcing a normative understanding. % Enabling critical (or preference-based) consumerism is an important aspect of expressing one's individuality. 

In line with these studies, we argue that providing a single meta yet multi-level label may be key to healthy and eco-friendly purchases. In this work, we present a visualisation which allows for personal prioritisation of nutrition or sustainability by providing both information in a multi-level manner. Yet, it offers meta-information at a single glance regarding a food's overall nutrition and environmental benefits. 

\section{Scale-Score}

\subsection{Requirements Gathering} \label{subsec:requirementsgathering}

Our goal is to design a label that shows nutritional and environmental information, and investigate its efficacy when it matters most, \textit{during} online shopping. To gather insights into the design strategy of the Scale-Score, we surveyed 120 German participants (80f; Appendix \ref{app:c}, Figure \ref{fig:age-stat}) and assessed their experience and knowledge towards Eco- and Nutri-Score (Subsection \ref{subsec:meta-labels}), online shopping, and queried their requirements towards a combined eco- and nutrition-label. Finally, we collected demographics data. All data is available online \cite{thesis}. 

The results show that most participants (95\,\%) are familiar with the Nutri-Score, having encountered it in German supermarkets during previous shopping trips. In contrast, only a third (36\,\%) stated they were familiar or understood the computation of the Nutri-Score. Regarding the Eco-Score, merely 7.5\,\% stated that they had already encountered this label in a German supermarket, with two participants (1.67\,\%) having a concrete understanding of how the Eco-Score is calculated. %It can be assumed that the Nutri-Score is much more widespread in Germany than the Eco-Score, which still seems to be in its infancy. 
Finally, the majority of participants (75\%) stated that the designs of both labels are understandable, suggesting the combining label use the same visual language. This echoes Sauve \textit{et al.}'s \cite{sauve2020econundrum} recommendations regarding a visualisation vocabulary (Subsection \ref{subsec:scale}) that is understood by different people or communities without being prescriptive, e.g. traffic light colour scheme.

Regarding participants' requirements towards a combined meta-score, almost half (45\,\%) stated they prioritise nutrition over sustainability, a quarter (27\,\%) assigned higher importance to sustainability over nutrition, and the remainder (28\,\%) want an equal ratio of 50:50 (Figure \ref{fig:prio-stat}, Appendix \ref{app:c}). 

In accordance with the literature, we encountered a dichotomy of requirements \cite{Bentvelzen, KIMURA2010_InteractiveEffectsCarbonFootprintInformation}: most participants (82\,\%) stated they prefer a simple label design which will give them a quick idea of a food's overall score (i.e. meta-label); yet, almost two thirds (62\,\%) wanted a lot of information on the joint label, going as far as different sustainability dimensions (i.e. multi-level labels). Finally, three quarters (71\,\%) stated they think it is important the final label design shows the individual Nutri- and Eco-Scores. Equipped with these insights, we designed the Scale-Score.

\subsection{Design}
\label{subsec:scale}

\begin{figure}[h!]%
    \centering
    \subfloat[\centering B-Level Scale-Score, with an A-level Nutri-Score, and a D-level Eco-Score. \label{fig:scale-ex}]{{\includegraphics[width=6cm]{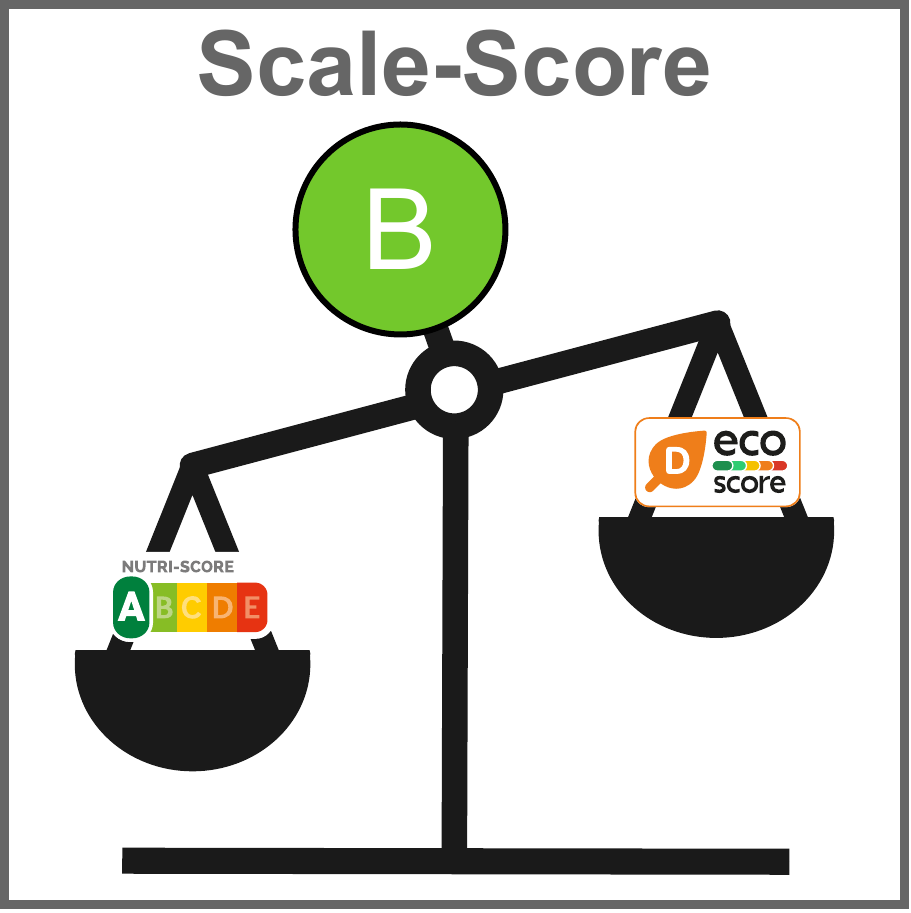} }}%
    \qquad
    \subfloat[{\centering Calculation of the Scale-Score, based on Nutri- and Eco-Score. \label{fig:example}}]{{\includegraphics[width=8cm]{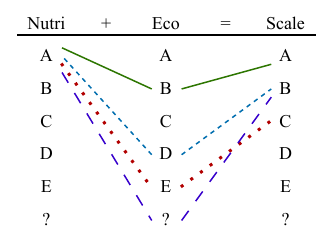}}}% 
    %former sketch1672852649300.png
    \caption{Design and calculation of the Scale-Score.}%
    \label{fig:scale-design-calculation}
    \Description{Figure 3 labelled `(a)' shows an exemplary B-level Scale-Score, composed of the individual Nutri-Score A-level and Eco-Score D-level rating of a product. Figure 3 labelled `(b)' shows further examples for calculating a product's Scale-Score. The examples given are: 1. An A-level Nutri-Score and a B-level Eco-Score result in an A-level Scale-Score. 2. An A-level Nutri-Score and a D-level Eco-Score result in a B-level Scale-Score. 3. An A-level Nutri-Score and an E-level Eco-Score result in a C-level Scale-Score. 4. An A-level Nutri-Score and an unknown Eco-Score result in a B-level Scale-Score.}
\end{figure}

The Scale-Score (Figure \ref{fig:scale-ex}) combines Nutri- and Eco-Scores into a single label. It functions as a \textit{multi-level label}, showing both Nutri- and Eco-Score are a part of the visualisation. The two labels are visually weighed against each other, which is represented by a classic beam scale. The Scale-Score is also a \textit{meta-label} as it shows an overall rating based on the product's Nutri- and Eco-Scores. The weighing pan of the Nutri-Score is deliberately shown hanging lower, since this label is weighted a little more heavily within the Scale-Score, as per user requirements (Subsection \ref{subsec:requirementsgathering}). 

The weight of the labels is designed in a very simplified way. As both, the Nutri- and the Eco-Score are five-colour scales, we computed their mean value, and in case of an uneven result, opted to go in favour of the Nutri-Score, prioritising nutrition over sustainability, in accordance with user requirements. For example, given a Nutri-Score of A and an Eco-Score with D, the mean score would be something between B and C; prioritising nutrition however, results in a Scale-Score of B (more examples in Figure \ref{fig:example}). In case a product's Eco- or Nutri-Score was missing, we reduced the non-missing score by one instead, respectively. If neither score was given, the final Scale-Score was marked with a question mark. The design of the Scale-Score was finalised after iteration with 3 pilot participants. 

\section{User Study}

\subsection{\textit{Groceries}: Online Shopping Environment}

To measure the efficacy of the Scale-Score quickly and efficiently, we created the mock online store \textit{Groceries} \ref{fig:all}. We shortly describe its implementation as the tool is available for other researchers to use and modify \cite{thesis} and GIT. The online store is implemented using the Apache XAMPP (v.~8.1.6) webserver including XAMPP Control Panel (v.~3.3.0). The client communicates via HTTP with the Apache server. The PHP engine is the heart of the operation: it manages the database, user profiles, and research methodology (e.g. in-between, Latin-Square). The database (formatted Open Food Facts database \cite{off}) contains all the products of the online store and predefined user accounts to allow individual purchase histories. Figure \ref{fig:xampp}, Appendix \ref{app:e} illustrate the software architecture of the online store. When conducting user research with \textit{Groceries}, prior to the shopping task, \textit{Groceries} displays the information sheet and asks for user consent. We tested \textit{Groceries}' usability and appropriateness for research in a pilot study. 

\subsection{Methods}
\label{subsec:study}

The study tested the impact of presenting the Scale-Score on the nutritional quality and environmental impact of the consumers' food choices, compared to the effects of both Nutri-Score and Eco-Score labels, and no persuasive technology. Thus, we used a within-subjects design with a single independent variable, visualisation, and three factors, Scale-Score, Nutri-/Eco-Score, and baseline with no visualisation. Dependent variables (i.e., shopping behaviour) were: 1) average environmental value of chosen products (based on Eco-Score calculations) and 2) average nutrition value of chosen products (based on Nutri-Score calculations). 

\begin{figure}[h!]
\centering
\resizebox{\columnwidth}{!}{%
\begin{tikzpicture}[background rectangle/.style={fill=gray!30}, show background rectangle] %optional background color
    % draw horizontal line
    \draw (1,0) -- (7,0);
    \draw (9,0) -- (13,0);

    % draw vertical lines
    \foreach \x in {1,3,5,7,9,11,13}
      %\draw (\x cm,3pt) -- (\x cm,-3pt);
      \filldraw (\x,0) circle (3pt);skills

	% draw nodes
	\scriptsize{ %or maybe \tiny
   	 
   	 %\draw (0,0); %begin of timeline
   	 
   	 \draw (1,0) node[above=5pt] {Sampling};
   	 \draw (3,0) node[below=5pt] {Introduction};
   	 \draw (5,0) node[above=5pt] {Pre-Study Q};
   	 \draw (7,0) node[below=5pt] {Condition};
   	 \draw (9,0) node[above=5pt] {Post-Condition Q};
   	 \draw (11,0) node[below=5pt] {Post-Study Q};
          \draw (13,0) node[above=5pt] {Debrief};

          %\draw[->,-stealth] (0,0) -- (0.9,0);
          \draw[->,-stealth] (1,0) -- (2.9,0);
          \draw[->,-stealth] (3,0) -- (4.9,0);
          \draw[->,-stealth] (5,0) -- (6.9,0);
          \draw[->,-stealth] (7,0) to [bend left=20] (8.9,0);
          \draw[->,-stealth] (9,0) to [bend left=20] (7.1,0);
          \draw[->,-stealth] (9,0) -- (10.9,0);
          \draw[->,-stealth] (11,0) -- (12.9,0);
   	 %\draw[->,-stealth] (13,0) -- (13.9,0); %end of timeline
   	 }
\end{tikzpicture}%
}
\caption{Experimental procedure. The abbreviation Q stands for questionnaire.}
\label{fig:timeline}
\Description{Figure 4 shows all steps of the experimental procedure, as they were already described in the text.}
\end{figure}
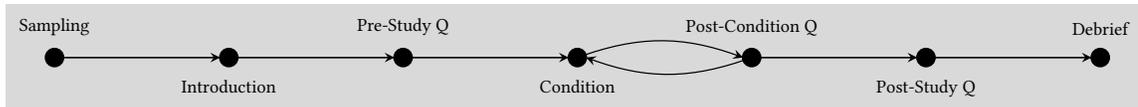

Following a brief, participants filled in a user behaviour and demographics questionnaire (Pre-Study, Figure \ref{fig:timeline}, \cite{thesis}). In each condition, participants shopped according to a shopping list with three items (cereal, milk, and peanut butter). Throughout, participants' screen was video recorded. Employing the think-aloud method, participants purchased foods and described their impressions and thoughts regarding the visualisations and the overall system. After each trial, they filled in a Post-Condition questionnaire. At the end of the study, we asked them to fill in two short Post-Study questionnaires; 1) level of support provided by the visualisations, and 2) label usability and system usability (Figure \ref{fig:timeline}). 

\subsection{Limitations}

As shopping task in each condition participants purchased one of each, peanut butter, cereals, and milk (12 random, non-repeating products per category e.g. to minimise brand-loyalty). We decided on these food categories as they have less `bad press' regarding sustainability and health compared to meats etc. Additionally, the study was conducted with the experimenter present to gather any additional insights regarding \textit{Groceries}' usability.

\subsection{Participants}

We recruited 12 participants (5f, $\mu =$ 39 years, $\sigma =$ 22.9 years) through our institution's forums. %, with 2/3 being between 20--29 years. 
The study lasted one hour. Participants were entered into a random draw to win their shopping basket, as an incentive to encourage normal purchase decisions. 

Regarding demographics, about 42\% reported their income situation as average and another 42\% as below average. 83.3\% are responsible for at least half of the grocery shopping in their household. All participants have experience in online shopping, 42\% with online grocery shopping. All participants stated familiarity with the Nutri-Score having seen it prior; the Eco-Score was seen by 16.7\,\% prior. 66.7\% do not follow any specific dietary regime. 16.7\% follow a vegetarian diet, giving taste, health, animal welfare, and lower price as their reasons. None had any food intolerances. 

\section{Results}

\subsection{Quantitative Results}

We used a one-way ANOVA with post-hoc Tukey-tests via IBM SPSS Statistics (v. 28.0.1.0) as statistic software. 
For this study an alpha ($\alpha$) of 0.05 was used. %An unknown Nutri-Score of products was calculated with the mean value 7; an unknown Eco-Score with the mean value 50. 
All data is available online (\cite{thesis}).

There was no statistically significant difference between the conditions on `nutrition' as determined by one-way ANOVA $(F(2.35) = 0.8 p = 0.458)$. However, a Tukey-post-hoc-test revealed that Scale- and Nutri-/Eco-Score means did not differ from each other ($p$-value $=0.994$), but both Nutri-/Eco-Score ($p$-value $=0.557$) and Scale-Score ($p$-value $=0.496$) differed compared to no score. Looking at the descriptive statistics, Scale-Score resulted in lower nutrition values (mean $=2.89$) compared to Nutri-/Eco-Score (mean $= 3.06$) and no score (mean $= 4.78$).

There also was no statistically significant difference between the conditions on `sustainability' as determined by one-way ANOVA $(F(2.35) = 1,301 p = 0.286)$. Tukey-post-hoc-test revealed a non-significant difference between Nutri-/Eco-Score ($p$-value $=0.595$) and Scale-Score ($p$-value $=0.810$) compared to no score. Looking at the descriptive statistics, Scale-Score resulted in lowest sustainability values (mean $=53.11$), Nutri-/Eco-Score in highest (mean $=59.78$) and no score in intermediate (mean $=55.69$).

\subsection{Qualitative Results}

Regarding sustainable and healthy decision making, half of the participants stated, that the visualisations impacted their choices most, whereas a third stated that previous purchasing habits impact their decisions most. 

Regarding usability, 75\% of participants strongly agreed with `The system was easy to use', 25\% mostly agreed. 41.7\% strongly disagreed with `The visualisation overtaxes me', and 41.7\% mostly disagreed with this statement. However, in accordance with previous literature and our previously conducted survey, participants (25\%) wished for more and customisable data to explain the scores. 

\section{Discussion}
% 
% Steps for a good discussion
% 
% 1) Summary: Start your discussion by reiterating the research problem, and consicely summarise your major findings. 
% 2) Interpretations: Identify patterns, discuss whether results meet expectations / support hypotheses, contextualise findings with related work / current literature, explain unexpected results, and offer potential alternative explanations. 
% 3) Implications: Discuss the meaning of your results for the scientific community; relate your work back to the literature. 
% 4) Discuss whether there are any practical implications? 
% 5) Limitations: Evaluate how limitations impacted your results, explain why your results are still valid. 
% 6) Recommendations: Make recommendations for practical implementation or further research.
% 

To achieve a global and successful transition to healthy and sustainable diets, systems and tools are needed to support consumers in this. We designed Scale-Score, a label that displays nutritional and environmental information. We compared Scale-Score's ability to support sustainable and nutritious online grocery shopping, compared to Nutri- and Eco-Score, and no guiding technology. 

We did not find significant differences in support provision of either visualisation compared to baseline, however there is a trend showing the Nutri- and Eco-Score combination may support consumers in sustainable and healthy decision making. The Scale-Score may support nutritious choices compared to no visualisation, however, worsened environmental impact of the basket compared to baseline. First, this may be due to the make-up of the Scale-Score: nutritional aspect weighted more into the final score. Consequently, a product that is marked with a good Scale-Score rating (e.g. B) may well contain an environmental `D' rating. As a result, the average sustainability score was worse compared to Nutri- and Eco-Score representation, resulting in Scale-Score's poor environmental performance. Second, participants may have ignored the multi-level information provided, given the small sizes of Nutri- and Eco-Score labels within the Scale-Score, contributing further to the de-valuation of environmental information. Finally, half of our participants stated they based their decisions on the visualisations, having potentially led to overwriting of prior sustainability knowledge. This suggests, that normative meta-labels may have a non-negligable impact on decision making, suggesting their continued use. In future work, we will weigh both data equally into the final meta-score and improve Scale-Score's visualisation such that it supports critical consumerism. 

Throughout this work, participants requested more and detailed information to accompany any given label. Unsurprisingly, yet surprisingly enlightening, research has shown nutrition labels have higher viewership given the `meaning of the label' is present \cite{GRAHAM2015nutritionfacts} --- that is, the Nutrition Facts panel. The Nutrition Facts panel is a label which is on the back of most packaging, showing what nutrients and other ingredients (to limit and get enough of) are in the food, whilst also giving daily recommendations for consumption. Given recently published works on consumer struggles to reflect over data \cite{Bentvelzen2022reflection} and the direction label research is heading in, namely purely visualisations, we assumed that visualisations suffice to support preference based decision making, leaving our `technical heritage' \cite{Feenberg1990CriticalTheoryfoTechnology} unquestioned. Yet, we found the Scale-Score to potentially disable critical consumerism, overwrite prior knowledge, and leaving consumers wanting for more information, particularly, sustainability. Therefore, we strongly believe that for any meta-label to be effective, for critical consumerism to be successful, we need to give consumers what they want: more information. Interestingly, we do precisely that since three decades through the Nutrition Facts panel. With this discussion, we call on all researchers to remember one critical aspect, often lacking in experimental, reductive research practices, context, and to invite more reflexivity into their practices. In future, we plan to conduct a study investigating the effects of a `Sustainability Facts' panel in combination with the Scale-Score, supporting true critical consumerism. 

% Finally, Nutri- and Eco-Score are (well) known labels in West-Europe, having been encountered by most participants. As peanut butter was an unfamiliar product for 1/3 of our participants, they may were forced to rely on labels in this category to tell if a product is sustainable or more nutritious. 

Finally, neither Nutri- nor Eco-Score are free from criticism. Both scores provide comparison between foods \textit{within} a category, not objective information whether the food is `healthy' or `sustainable'. The Eco-Score particularly is under scrutiny as its Life Cycle Assessment takes criteria into account that are criticised for accelerating environmental degradation (e.g. prioritises intensive agriculture) \cite{FOPfrance}, and for greenwashing practices \cite{LeMondeGreenWashing, LeMondeGreenWashing2}. In future, we will utilise the Planet Score \cite{PlanetScore} instead, which was developed by consumer- and environmental protection associations, instead of the industrial complex, as is the Eco-Score \cite{BNN}. We will do this switch as we strongly believe that researchers have an obligation to support and disseminate best practices to ensure a sustainable future, especially Sustainable HCI researchers. 

Reflecting on the results, the Scale-Score is a promising start into user-preference based food labels to support nutritious and sustainable diets, in a meta yet multi-level manner. This work demonstrated and argues that there is a dire necessity for more user-centred food label research, embedded in personal informatics and reflexivity. 

\section{Conclusions}
% 
% Steps for a good conclusion
% 
% 1) Restate the problem
% 2) Sum up the paper
% 3) Discuss the implications
% 

In this paper we present Scale-Score, a meta yet multi-level label which comprises of both, nutritional and environmental information. We describe the design of Scale-Score, the implementation of the open-source mock online grocery environment, and an experiment investigating its support during online grocery shopping. Our findings show that the design of the Scale-Score supports nutritious but not sustainable decision making. Our results and discussion on meta and multi-level aspects of nutrition and sustainability food labels may alter the label design space, making it more user-centred, reflexive, and ultimately, more sustainable. 

% \section{SIGCHI Extended Abstracts}

% The ``\verb|sigchi-a|'' template style (available only in \LaTeX\ and
% not in Word) produces a landscape-orientation formatted article, with
% a wide left margin. Three environments are available for use with the
% ``\verb|sigchi-a|'' template style, and produce formatted output in
% the margin:
% \begin{description}
% \item[\texttt{sidebar}:]  Place formatted text in the margin.
% \item[\texttt{marginfigure}:] Place a figure in the margin.
% \item[\texttt{margintable}:] Place a table in the margin.
% \end{description}

%%
%% The acknowledgements section is defined using the "acks" environment
% \begin{acks}
% This study was not funded. No conflicts of interests are declared by the authors. We would like to thank all our colleagues at \censor{Carl von Ossietzky University of Oldenburg} and at the \censor{OFFIS Institute for Informatics} for their unwavering support and feedback. Finally, we would like to thank \censor{family and friends of first author?}.
% \end{acks}

\bibliographystyle{ACM-Reference-Format}
%\bibliography{sample-base}
\bibliography{main}


\begin{thebibliography}{63}

%%% the Web address.

\ifx \showCODEN    \undefined \def \showCODEN     #1{\unskip}     \fi
\ifx \showDOI      \undefined \def \showDOI       #1{#1}\fi
\ifx \showISBNx    \undefined \def \showISBNx     #1{\unskip}     \fi
\ifx \showISBNxiii \undefined \def \showISBNxiii  #1{\unskip}     \fi
\ifx \showISSN     \undefined \def \showISSN      #1{\unskip}     \fi
\ifx \showLCCN     \undefined \def \showLCCN      #1{\unskip}     \fi
\ifx \shownote     \undefined \def \shownote      #1{#1}          \fi
\ifx \showarticletitle \undefined \def \showarticletitle #1{#1}   \fi
\ifx \showURL      \undefined \def \showURL       {\relax}        \fi
% invisible to TeX
\providecommand\bibfield[2]{#2}
\providecommand\bibinfo[2]{#2}
\providecommand\natexlab[1]{#1}
\providecommand\showeprint[2][]{arXiv:#2}

\bibitem[Apostolidis and McLeay(2019)]%
        {APOSTOLIDIS2019109}
\bibfield{author}{\bibinfo{person}{Chrysostomos Apostolidis} {and}
  \bibinfo{person}{Fraser McLeay}.} \bibinfo{year}{2019}\natexlab{}.
\newblock \showarticletitle{To meat or not to meat? Comparing empowered meat
  consumers’ and anti-consumers’ preferences for sustainability labels}.
\newblock \bibinfo{journal}{\emph{Food Quality and Preference}}
  \bibinfo{volume}{77} (\bibinfo{year}{2019}), \bibinfo{pages}{109--122}.
\newblock
\showISSN{0950-3293}
\urldef\tempurl%
\url{https://doi.org/10.1016/j.foodqual.2019.04.008}
\showDOI{\tempurl}


\bibitem[Asioli et~al\mbox{.}(2020)]%
        {asioli2020sustainability}
\bibfield{author}{\bibinfo{person}{Daniele Asioli}, \bibinfo{person}{Jessica
  Aschemann-Witzel}, {and} \bibinfo{person}{Rodolfo~M Nayga~Jr}.}
  \bibinfo{year}{2020}\natexlab{}.
\newblock \showarticletitle{Sustainability-related food labels}.
\newblock \bibinfo{journal}{\emph{Annual Review of Resource Economics}}
  \bibinfo{volume}{12} (\bibinfo{year}{2020}), \bibinfo{pages}{171--185}.
\newblock


\bibitem[Bentvelzen et~al\mbox{.}(2021)]%
        {Bentvelzen}
\bibfield{author}{\bibinfo{person}{Marit Bentvelzen}, \bibinfo{person}{Jasmin
  Niess}, {and} \bibinfo{person}{Pawe\l{}~W. Wo\'{z}niak}.}
  \bibinfo{year}{2021}\natexlab{}.
\newblock \showarticletitle{The Technology-Mediated Reflection Model: Barriers
  and Assistance in Data-Driven Reflection}. In
  \bibinfo{booktitle}{\emph{Proceedings of the 2021 CHI Conference on Human
  Factors in Computing Systems}} (Yokohama, Japan) \emph{(\bibinfo{series}{CHI
  '21})}. \bibinfo{publisher}{Association for Computing Machinery},
  \bibinfo{address}{New York, NY, USA}, Article \bibinfo{articleno}{246},
  \bibinfo{numpages}{12}~pages.
\newblock
\showISBNx{9781450380966}
\urldef\tempurl%
\url{https://doi.org/10.1145/3411764.3445505}
\showDOI{\tempurl}


\bibitem[Bentvelzen et~al\mbox{.}(2022)]%
        {Bentvelzen2022reflection}
\bibfield{author}{\bibinfo{person}{Marit Bentvelzen},
  \bibinfo{person}{Pawe\l{}~W. Wo\'{z}niak}, \bibinfo{person}{Pia~S.F. Herbes},
  \bibinfo{person}{Evropi Stefanidi}, {and} \bibinfo{person}{Jasmin Niess}.}
  \bibinfo{year}{2022}\natexlab{}.
\newblock \showarticletitle{Revisiting Reflection in HCI: Four Design Resources
  for Technologies That Support Reflection}.
\newblock \bibinfo{journal}{\emph{Proc. ACM Interact. Mob. Wearable Ubiquitous
  Technol.}} \bibinfo{volume}{6}, \bibinfo{number}{1}, Article
  \bibinfo{articleno}{2} (\bibinfo{date}{mar} \bibinfo{year}{2022}),
  \bibinfo{numpages}{27}~pages.
\newblock
\urldef\tempurl%
\url{https://doi.org/10.1145/3517233}
\showDOI{\tempurl}


\bibitem[{Burkhard Schilz}(2017)]%
        {xampp}
\bibfield{author}{\bibinfo{person}{{Burkhard Schilz}}.}
  \bibinfo{year}{2017}\natexlab{}.
\newblock \bibinfo{booktitle}{\emph{Webserver Apache \& Xampp}}.
\newblock
\urldef\tempurl%
\url{https://slideplayer.org/slide/3686064/}
\showURL{%
\tempurl}


\bibitem[Carling et~al\mbox{.}(2015)]%
        {CARLING2015onlinetransportCO2}
\bibfield{author}{\bibinfo{person}{Kenneth Carling}, \bibinfo{person}{Mengjie
  Han}, \bibinfo{person}{Johan Håkansson}, \bibinfo{person}{Xiangli Meng},
  {and} \bibinfo{person}{Niklas Rudholm}.} \bibinfo{year}{2015}\natexlab{}.
\newblock \showarticletitle{Measuring transport related CO2 emissions induced
  by online and brick-and-mortar retailing}.
\newblock \bibinfo{journal}{\emph{Transportation Research Part D: Transport and
  Environment}}  \bibinfo{volume}{40} (\bibinfo{year}{2015}),
  \bibinfo{pages}{28--42}.
\newblock
\showISSN{1361-9209}
\urldef\tempurl%
\url{https://doi.org/10.1016/j.trd.2015.07.010}
\showDOI{\tempurl}


\bibitem[Clear and Friday(2012)]%
        {clear2012designing}
\bibfield{author}{\bibinfo{person}{Adrian Clear} {and} \bibinfo{person}{Adrian
  Friday}.} \bibinfo{year}{2012}\natexlab{}.
\newblock \showarticletitle{Designing a Food'Qualculator'}. In
  \bibinfo{booktitle}{\emph{DIS 2012 workshop on Food for Thought: Designing
  for Critical Reflection on Food Practices. Newcastle, UK}}.
\newblock


\bibitem[{De Temmerman} et~al\mbox{.}(2021)]%
        {DETEMMERMAN2021104995}
\bibfield{author}{\bibinfo{person}{Joyce {De Temmerman}}, \bibinfo{person}{Eva
  Heeremans}, \bibinfo{person}{Hendrik Slabbinck}, {and} \bibinfo{person}{Iris
  Vermeir}.} \bibinfo{year}{2021}\natexlab{}.
\newblock \showarticletitle{The impact of the Nutri-Score nutrition label on
  perceived healthiness and purchase intentions}.
\newblock \bibinfo{journal}{\emph{Appetite}}  \bibinfo{volume}{157}
  (\bibinfo{year}{2021}), \bibinfo{pages}{104995}.
\newblock
\showISSN{0195-6663}
\urldef\tempurl%
\url{https://doi.org/10.1016/j.appet.2020.104995}
\showDOI{\tempurl}


\bibitem[Dendler(2014)]%
        {dendler2014metalabelling}
\bibfield{author}{\bibinfo{person}{Leonie Dendler}.}
  \bibinfo{year}{2014}\natexlab{}.
\newblock \showarticletitle{Sustainability Meta Labelling: an effective measure
  to facilitate more sustainable consumption and production?}
\newblock \bibinfo{journal}{\emph{Journal of Cleaner Production}}
  \bibinfo{volume}{63} (\bibinfo{year}{2014}), \bibinfo{pages}{74--83}.
\newblock
\showISSN{0959-6526}
\urldef\tempurl%
\url{https://doi.org/10.1016/j.jclepro.2013.04.037}
\showDOI{\tempurl}
\newblock
\shownote{Special Volume: Sustainable Production, Consumption and Livelihoods:
  Global and Regional Research Perspectives}.


\bibitem[Dinu et~al\mbox{.}(2017)]%
        {dinu2017healthyvegan}
\bibfield{author}{\bibinfo{person}{Monica Dinu}, \bibinfo{person}{Rosanna
  Abbate}, \bibinfo{person}{Gian~Franco Gensini}, \bibinfo{person}{Alessandro
  Casini}, {and} \bibinfo{person}{Francesco Sofi}.}
  \bibinfo{year}{2017}\natexlab{}.
\newblock \showarticletitle{Vegetarian, vegan diets and multiple health
  outcomes: A systematic review with meta-analysis of observational studies}.
\newblock \bibinfo{journal}{\emph{Critical Reviews in Food Science and
  Nutrition}} \bibinfo{volume}{57}, \bibinfo{number}{17}
  (\bibinfo{year}{2017}), \bibinfo{pages}{3640--3649}.
\newblock
\urldef\tempurl%
\url{https://doi.org/10.1080/10408398.2016.1138447}
\showDOI{\tempurl}
\showeprint{https://doi.org/10.1080/10408398.2016.1138447}
\newblock
\shownote{PMID: 26853923}.


\bibitem[Druschba(2022)]%
        {thesis}
\bibfield{author}{\bibinfo{person}{Marco Druschba}.}
  \bibinfo{year}{2022}\natexlab{}.
\newblock \bibinfo{booktitle}{\emph{Entwicklung eines Online-Shops zur Erhebung
  der Auswirkungen verschiedener Lebensmittelkennzeichnungen auf das
  nachhaltigkeits- und nährwerttechnische Einkaufsverhalten bei
  Online-Einkäufen}}.
\newblock
\urldef\tempurl%
\url{https://oops.uni-oldenburg.de/id/eprint/5512}
\showURL{%
\tempurl}


\bibitem[Dufeu et~al\mbox{.}(2014)]%
        {Dufeu2014multilabellingandWTP}
\bibfield{author}{\bibinfo{person}{Ivan Dufeu}, \bibinfo{person}{Jean-Marc
  Ferrandi}, \bibinfo{person}{Patrick Gabriel}, {and}
  \bibinfo{person}{Marine~Le Gall-Ely}.} \bibinfo{year}{2014}\natexlab{}.
\newblock \showarticletitle{Socio-environmental multi-labelling and consumer
  willingness to pay}.
\newblock \bibinfo{journal}{\emph{Recherche et Applications en Marketing
  (English Edition)}} \bibinfo{volume}{29}, \bibinfo{number}{3}
  (\bibinfo{year}{2014}), \bibinfo{pages}{35--56}.
\newblock
\urldef\tempurl%
\url{https://doi.org/10.1177/2051570714542063}
\showDOI{\tempurl}
\showeprint{https://doi.org/10.1177/2051570714542063}


\bibitem[Egnell et~al\mbox{.}(2018)]%
        {egnell2018objective}
\bibfield{author}{\bibinfo{person}{Manon Egnell}, \bibinfo{person}{Pauline
  Ducrot}, \bibinfo{person}{Mathilde Touvier}, \bibinfo{person}{Benjamin
  All{\`e}s}, \bibinfo{person}{Serge Hercberg}, \bibinfo{person}{Emmanuelle
  Kesse-Guyot}, {and} \bibinfo{person}{Chantal Julia}.}
  \bibinfo{year}{2018}\natexlab{}.
\newblock \showarticletitle{Objective understanding of Nutri-Score
  Front-Of-Package nutrition label according to individual characteristics of
  subjects: Comparisons with other format labels}.
\newblock \bibinfo{journal}{\emph{PloS one}} \bibinfo{volume}{13},
  \bibinfo{number}{8} (\bibinfo{year}{2018}), \bibinfo{pages}{e0202095}.
\newblock
\urldef\tempurl%
\url{https://doi.org/10.1371/journal.pone.0202095}
\showDOI{\tempurl}


\bibitem[et~Mathilde~Gérard(2020)]%
        {LeMondeGreenWashing}
\bibfield{author}{\bibinfo{person}{Stéphane~Foucart et Mathilde~Gérard}.}
  \bibinfo{year}{2020}\natexlab{}.
\newblock \bibinfo{booktitle}{\emph{La certification agricole « HVE » sous le
  feu d’une nouvelle critique}}.
\newblock
\urldef\tempurl%
\url{https://www.lemonde.fr/planete/article/2021/05/25/la-certification-agricole-hve-sous-le-feu-d-une-nouvelle-critique_6081444_3244.html}
\showURL{%
\tempurl}


\bibitem[Facts(2022)]%
        {eslogo}
\bibfield{author}{\bibinfo{person}{Open~Food Facts}.}
  \bibinfo{year}{2022}\natexlab{}.
\newblock \bibinfo{booktitle}{\emph{Eco-Score: the environmental impact of food
  products}}.
\newblock
\urldef\tempurl%
\url{https://de-en.openfoodfacts.org/eco-score-the-environmental-impact-of-food-products}
\showURL{%
\tempurl}


\bibitem[Facts(2023)]%
        {off}
\bibfield{author}{\bibinfo{person}{Open~Food Facts}.}
  \bibinfo{year}{2023}\natexlab{}.
\newblock \bibinfo{booktitle}{\emph{Open Food Facts -- World}}.
\newblock
\urldef\tempurl%
\url{https://world.openfoodfacts.org/}
\showURL{%
\tempurl}


\bibitem[Feenberg(1990)]%
        {Feenberg1990CriticalTheoryfoTechnology}
\bibfield{author}{\bibinfo{person}{Andrew Feenberg}.}
  \bibinfo{year}{1990}\natexlab{}.
\newblock \showarticletitle{The critical theory of technology}.
\newblock \bibinfo{journal}{\emph{Capitalism Nature Socialism}}
  \bibinfo{volume}{1}, \bibinfo{number}{5} (\bibinfo{year}{1990}),
  \bibinfo{pages}{17--45}.
\newblock
\urldef\tempurl%
\url{https://doi.org/10.1080/10455759009358413}
\showDOI{\tempurl}
\showeprint{https://doi.org/10.1080/10455759009358413}


\bibitem[for National~Statistics(2022)]%
        {InternetStatistics}
\bibfield{author}{\bibinfo{person}{Office for National~Statistics}.}
  \bibinfo{year}{2022}\natexlab{}.
\newblock \bibinfo{booktitle}{\emph{Internet sales as a percentage of total
  retail sales (ratio) (\%)}}.
\newblock
\urldef\tempurl%
\url{https://www.ons.gov.uk/businessindustryandtrade/retailindustry/timeseries/j4mc/drsi}
\showURL{%
\tempurl}


\bibitem[Froehlich et~al\mbox{.}(2010)]%
        {Froehlich2010ecotechnology}
\bibfield{author}{\bibinfo{person}{Jon Froehlich}, \bibinfo{person}{Leah
  Findlater}, {and} \bibinfo{person}{James Landay}.}
  \bibinfo{year}{2010}\natexlab{}.
\newblock \showarticletitle{The Design of Eco-Feedback Technology}. In
  \bibinfo{booktitle}{\emph{Proceedings of the SIGCHI Conference on Human
  Factors in Computing Systems}} (Atlanta, Georgia, USA)
  \emph{(\bibinfo{series}{CHI '10})}. \bibinfo{publisher}{Association for
  Computing Machinery}, \bibinfo{address}{New York, NY, USA},
  \bibinfo{pages}{1999–2008}.
\newblock
\showISBNx{9781605589299}
\urldef\tempurl%
\url{https://doi.org/10.1145/1753326.1753629}
\showDOI{\tempurl}


\bibitem[Fuchs et~al\mbox{.}(2022)]%
        {fuchs2022betterchoice}
\bibfield{author}{\bibinfo{person}{Klaus~L. Fuchs}, \bibinfo{person}{Jie Lian},
  \bibinfo{person}{Leonard Michels}, \bibinfo{person}{Simon Mayer},
  \bibinfo{person}{Enrico Toniato}, {and} \bibinfo{person}{Verena Tiefenbeck}.}
  \bibinfo{year}{2022}\natexlab{}.
\newblock \showarticletitle{Effects of Digital Food Labels on Healthy Food
  Choices in Online Grocery Shopping}.
\newblock \bibinfo{journal}{\emph{Nutrients}} \bibinfo{volume}{14},
  \bibinfo{number}{10} (\bibinfo{year}{2022}).
\newblock
\showISSN{2072-6643}
\urldef\tempurl%
\url{https://doi.org/10.3390/nu14102044}
\showDOI{\tempurl}


\bibitem[für Ernährung~und Landwirtschaft(2019)]%
        {BMLnutriscore}
\bibfield{author}{\bibinfo{person}{Bundesministerium für Ernährung~und
  Landwirtschaft}.} \bibinfo{year}{2019}\natexlab{}.
\newblock \bibinfo{booktitle}{\emph{Evaluation von erweiterten
  Nährwertkennzeichnungs-Modellen – Ergebnisbericht der
  Fokusgruppendiskussionen}}.
\newblock
\urldef\tempurl%
\url{https://www.test-achats.be/sante/alimentation-et-nutrition/alimentation-saine/presse/2020/sondage-de-test-achats-une-majorite-est-favorable-a-un-nutri-score-obligatoire}
\showURL{%
\tempurl}


\bibitem[Graham et~al\mbox{.}(2015)]%
        {GRAHAM2015nutritionfacts}
\bibfield{author}{\bibinfo{person}{Dan.~J. Graham}, \bibinfo{person}{Charles
  Heidrick}, {and} \bibinfo{person}{Katie Hodgin}.}
  \bibinfo{year}{2015}\natexlab{}.
\newblock \showarticletitle{Nutrition Label Viewing during a Food-Selection
  Task: Front-of-Package Labels vs Nutrition Facts Labels}.
\newblock \bibinfo{journal}{\emph{Journal of the Academy of Nutrition and
  Dietetics}} \bibinfo{volume}{115}, \bibinfo{number}{10}
  (\bibinfo{year}{2015}), \bibinfo{pages}{1636--1646}.
\newblock
\showISSN{2212-2672}
\urldef\tempurl%
\url{https://doi.org/10.1016/j.jand.2015.02.019}
\showDOI{\tempurl}


\bibitem[Grebitus et~al\mbox{.}(2015)]%
        {GrebitusRoosenSeitz+2015+73+81}
\bibfield{author}{\bibinfo{person}{Carola Grebitus}, \bibinfo{person}{Jutta
  Roosen}, {and} \bibinfo{person}{Carolin~Claudia Seitz}.}
  \bibinfo{year}{2015}\natexlab{}.
\newblock \showarticletitle{Visual Attention and Choice: A Behavioral Economics
  Perspective on Food Decisions}.
\newblock \bibinfo{journal}{\emph{Journal of Agricultural \& Food Industrial
  Organization}} \bibinfo{volume}{13}, \bibinfo{number}{1}
  (\bibinfo{year}{2015}), \bibinfo{pages}{73--81}.
\newblock
\urldef\tempurl%
\url{https://doi.org/doi:10.1515/jafio-2015-0017}
\showDOI{\tempurl}


\bibitem[Hanss and Doran(2020)]%
        {Hanss2020_PerceivedConsumerEffectiveness}
\bibfield{author}{\bibinfo{person}{Daniel Hanss} {and} \bibinfo{person}{Rouven
  Doran}.} \bibinfo{year}{2020}\natexlab{}.
\newblock \showarticletitle{Perceived consumer effectiveness}.
\newblock \bibinfo{journal}{\emph{Responsible Consumption and Production}}
  (\bibinfo{year}{2020}), \bibinfo{pages}{535--544}.
\newblock


\bibitem[Harrington et~al\mbox{.}(2019)]%
        {Harrington2019Flicc}
\bibfield{author}{\bibinfo{person}{Richard~A Harrington},
  \bibinfo{person}{Peter Scarborough}, \bibinfo{person}{Charo Hodgkins},
  \bibinfo{person}{Monique~M Raats}, \bibinfo{person}{Gill Cowburn},
  \bibinfo{person}{Moira Dean}, \bibinfo{person}{Aiden Doherty},
  \bibinfo{person}{Charlie Foster}, \bibinfo{person}{Edmund Juszczak},
  \bibinfo{person}{Cliona Ni~Mhurchu}, \bibinfo{person}{Naomi Winstone},
  \bibinfo{person}{Richard Shepherd}, \bibinfo{person}{Lada Timotijevic}, {and}
  \bibinfo{person}{Mike Rayner}.} \bibinfo{year}{2019}\natexlab{}.
\newblock \showarticletitle{A Pilot Randomized Controlled Trial of a Digital
  Intervention Aimed at Improving Food Purchasing Behavior: The Front-of-Pack
  Food Labels Impact on Consumer Choice Study}.
\newblock \bibinfo{journal}{\emph{JMIR Form Res}} \bibinfo{volume}{3},
  \bibinfo{number}{2} (\bibinfo{date}{08 Apr} \bibinfo{year}{2019}),
  \bibinfo{pages}{e9910}.
\newblock
\showISSN{2561-326X}
\urldef\tempurl%
\url{https://doi.org/10.2196/formative.9910}
\showDOI{\tempurl}


\bibitem[Hekler et~al\mbox{.}(2013)]%
        {Hekler2013_MindTheTheoreticalGap}
\bibfield{author}{\bibinfo{person}{Eric~B. Hekler}, \bibinfo{person}{Predrag
  Klasnja}, \bibinfo{person}{Jon~E. Froehlich}, {and}
  \bibinfo{person}{Matthew~P. Buman}.} \bibinfo{year}{2013}\natexlab{}.
\newblock \showarticletitle{Mind the Theoretical Gap: Interpreting, Using, and
  Developing Behavioral Theory in HCI Research}. In
  \bibinfo{booktitle}{\emph{Proceedings of the SIGCHI Conference on Human
  Factors in Computing Systems}} (Paris, France) \emph{(\bibinfo{series}{CHI
  '13})}. \bibinfo{publisher}{Association for Computing Machinery},
  \bibinfo{address}{New York, NY, USA}, \bibinfo{pages}{3307–3316}.
\newblock
\showISBNx{9781450318990}
\urldef\tempurl%
\url{https://doi.org/10.1145/2470654.2466452}
\showDOI{\tempurl}


\bibitem[Hekler et~al\mbox{.}(2016)]%
        {Hekler2016_AgileScienceCreatingUsefulProducts}
\bibfield{author}{\bibinfo{person}{Eric~B. Hekler}, \bibinfo{person}{Predrag
  Klasnja}, \bibinfo{person}{William~T. Riley}, \bibinfo{person}{Matthew~P.
  Buman}, \bibinfo{person}{Jennifer Huberty}, \bibinfo{person}{Daniel~E.
  Rivera}, {and} \bibinfo{person}{Cesar~A. Martin}.}
  \bibinfo{year}{2016}\natexlab{}.
\newblock \showarticletitle{{Agile science: creating useful products for
  behavior change in the real world}}.
\newblock \bibinfo{journal}{\emph{Translational Behavioral Medicine}}
  \bibinfo{volume}{6}, \bibinfo{number}{2} (\bibinfo{date}{02}
  \bibinfo{year}{2016}), \bibinfo{pages}{317--328}.
\newblock
\showISSN{1869-6716}
\urldef\tempurl%
\url{https://doi.org/10.1007/s13142-016-0395-7}
\showDOI{\tempurl}
\showeprint{https://academic.oup.com/tbm/article-pdf/6/2/317/22066988/13142\_2016\_article\_395.pdf}


\bibitem[Hercberg et~al\mbox{.}(2021)]%
        {hercberg2021nutri}
\bibfield{author}{\bibinfo{person}{Serge Hercberg}, \bibinfo{person}{Mathilde
  Touvier}, \bibinfo{person}{Jordi Salas-Salvado}, {et~al\mbox{.}}}
  \bibinfo{year}{2021}\natexlab{}.
\newblock \showarticletitle{The Nutri-Score nutrition label}.
\newblock \bibinfo{journal}{\emph{International Journal for Vitamin and
  Nutrition Research}} (\bibinfo{year}{2021}), \bibinfo{pages}{1--11}.
\newblock
\urldef\tempurl%
\url{https://doi.org/10.1024/0300-9831/a000722}
\showDOI{\tempurl}


\bibitem[Index(2023)]%
        {EcolabelIndex}
\bibfield{author}{\bibinfo{person}{Ecolabel Index}.}
  \bibinfo{year}{2023}\natexlab{}.
\newblock \bibinfo{booktitle}{\emph{Ecolabel Index, the global directory of
  ecolabels}}.
\newblock
\urldef\tempurl%
\url{https://www.ecolabelindex.com/ecolabels/?search=food\&as_values_053=}
\showURL{%
\tempurl}


\bibitem[{Isabel Sonntag} et~al\mbox{.}(2023)]%
        {ISABELSONNTAG2023labeljungle}
\bibfield{author}{\bibinfo{person}{Winnie {Isabel Sonntag}},
  \bibinfo{person}{Dominic Lemken}, \bibinfo{person}{Achim Spiller}, {and}
  \bibinfo{person}{Maureen Schulze}.} \bibinfo{year}{2023}\natexlab{}.
\newblock \showarticletitle{Welcome to the (label) jungle? Analyzing how
  consumers deal with intra-sustainability label trade-offs on food}.
\newblock \bibinfo{journal}{\emph{Food Quality and Preference}}
  \bibinfo{volume}{104} (\bibinfo{year}{2023}), \bibinfo{pages}{104746}.
\newblock
\showISSN{0950-3293}
\urldef\tempurl%
\url{https://doi.org/10.1016/j.foodqual.2022.104746}
\showDOI{\tempurl}


\bibitem[Kaczorowska et~al\mbox{.}(2019)]%
        {kaczorowska2019perceivedproductvalue}
\bibfield{author}{\bibinfo{person}{Joanna Kaczorowska},
  \bibinfo{person}{Krystyna Rejman}, \bibinfo{person}{Ewa Halicka},
  \bibinfo{person}{Agata Szczebyło}, {and} \bibinfo{person}{Hanna
  Górska-Warsewicz}.} \bibinfo{year}{2019}\natexlab{}.
\newblock \showarticletitle{Impact of Food Sustainability Labels on the
  Perceived Product Value and Price Expectations of Urban Consumers}.
\newblock \bibinfo{journal}{\emph{Sustainability}} \bibinfo{volume}{11},
  \bibinfo{number}{24} (\bibinfo{year}{2019}).
\newblock
\showISSN{2071-1050}
\urldef\tempurl%
\url{https://doi.org/10.3390/su11247240}
\showDOI{\tempurl}


\bibitem[Kimura et~al\mbox{.}(2010)]%
        {KIMURA2010_InteractiveEffectsCarbonFootprintInformation}
\bibfield{author}{\bibinfo{person}{Atsushi Kimura}, \bibinfo{person}{Yuji
  Wada}, \bibinfo{person}{Akiko Kamada}, \bibinfo{person}{Tomohiro Masuda},
  \bibinfo{person}{Masako Okamoto}, \bibinfo{person}{Sho ichi Goto},
  \bibinfo{person}{Daisuke Tsuzuki}, \bibinfo{person}{Dongsheng Cai},
  \bibinfo{person}{Takashi Oka}, {and} \bibinfo{person}{Ippeita Dan}.}
  \bibinfo{year}{2010}\natexlab{}.
\newblock \showarticletitle{Interactive effects of carbon footprint information
  and its accessibility on value and subjective qualities of food products}.
\newblock \bibinfo{journal}{\emph{Appetite}} \bibinfo{volume}{55},
  \bibinfo{number}{2} (\bibinfo{year}{2010}), \bibinfo{pages}{271 -- 278}.
\newblock
\showISSN{0195-6663}
\urldef\tempurl%
\url{https://doi.org/10.1016/j.appet.2010.06.013}
\showDOI{\tempurl}


\bibitem[Lawo et~al\mbox{.}(2021)]%
        {lawo2021criticalconsumer}
\bibfield{author}{\bibinfo{person}{Dennis Lawo}, \bibinfo{person}{Thomas
  Neifer}, \bibinfo{person}{Margarita Esau}, {and} \bibinfo{person}{Gunnar
  Stevens}.} \bibinfo{year}{2021}\natexlab{}.
\newblock \showarticletitle{Buying the ‘Right’ Thing: Designing Food
  Recommender Systems with Critical Consumers}. In
  \bibinfo{booktitle}{\emph{Proceedings of the 2021 CHI Conference on Human
  Factors in Computing Systems}} (Yokohama, Japan) \emph{(\bibinfo{series}{CHI
  '21})}. \bibinfo{publisher}{Association for Computing Machinery},
  \bibinfo{address}{New York, NY, USA}, Article \bibinfo{articleno}{85},
  \bibinfo{numpages}{13}~pages.
\newblock
\showISBNx{9781450380966}
\urldef\tempurl%
\url{https://doi.org/10.1145/3411764.3445264}
\showDOI{\tempurl}


\bibitem[Lemken et~al\mbox{.}(2021)]%
        {lemken2021climatescorelabel}
\bibfield{author}{\bibinfo{person}{Dominic Lemken}, \bibinfo{person}{Anke
  Zühlsdorf}, {and} \bibinfo{person}{Achim Spiller}.}
  \bibinfo{year}{2021}\natexlab{}.
\newblock \showarticletitle{Improving Consumers’ Understanding and Use of
  Carbon Footprint Labels on Food: Proposal for a Climate Score Label}.
\newblock \bibinfo{journal}{\emph{EuroChoices}} \bibinfo{volume}{20},
  \bibinfo{number}{2} (\bibinfo{year}{2021}), \bibinfo{pages}{23--29}.
\newblock
\urldef\tempurl%
\url{https://doi.org/10.1111/1746-692X.12321}
\showDOI{\tempurl}
\showeprint{https://onlinelibrary.wiley.com/doi/pdf/10.1111/1746-692X.12321}


\bibitem[Lentz et~al\mbox{.}(2018)]%
        {LENTZ2018GaugingAttitudesMeatConsumption}
\bibfield{author}{\bibinfo{person}{Garrett Lentz}, \bibinfo{person}{Sean
  Connelly}, \bibinfo{person}{Miranda Mirosa}, {and} \bibinfo{person}{Tim
  Jowett}.} \bibinfo{year}{2018}\natexlab{}.
\newblock \showarticletitle{Gauging attitudes and behaviours: Meat consumption
  and potential reduction}.
\newblock \bibinfo{journal}{\emph{Appetite}}  \bibinfo{volume}{127}
  (\bibinfo{year}{2018}), \bibinfo{pages}{230 -- 241}.
\newblock
\showISSN{0195-6663}
\urldef\tempurl%
\url{https://doi.org/10.1016/j.appet.2018.04.015}
\showDOI{\tempurl}


\bibitem[Luchs and Kumar(2017)]%
        {luchs2017tradeoffs}
\bibfield{author}{\bibinfo{person}{Michael~G Luchs} {and} \bibinfo{person}{Minu
  Kumar}.} \bibinfo{year}{2017}\natexlab{}.
\newblock \showarticletitle{“Yes, but this other one looks better/works
  better”: How do consumers respond to trade-offs between sustainability and
  other valued attributes?}
\newblock \bibinfo{journal}{\emph{Journal of Business Ethics}}
  \bibinfo{volume}{140}, \bibinfo{number}{3} (\bibinfo{year}{2017}),
  \bibinfo{pages}{567--584}.
\newblock


\bibitem[Luo et~al\mbox{.}(2017)]%
        {Luo2017_OnlineEngagementWebbaseSupermarketHealthProgram}
\bibfield{author}{\bibinfo{person}{Ling Luo}, \bibinfo{person}{Bin Li},
  \bibinfo{person}{Shlomo Berkovsky}, \bibinfo{person}{Irena Koprinska}, {and}
  \bibinfo{person}{Fang Chen}.} \bibinfo{year}{2017}\natexlab{}.
\newblock \showarticletitle{Online Engagement for a Healthier You: A Case Study
  of Web-Based Supermarket Health Program}. In
  \bibinfo{booktitle}{\emph{Proceedings of the 26th International Conference on
  World Wide Web Companion}} (Perth, Australia) \emph{(\bibinfo{series}{WWW '17
  Companion})}. \bibinfo{publisher}{International World Wide Web Conferences
  Steering Committee}, \bibinfo{address}{Republic and Canton of Geneva, CHE},
  \bibinfo{pages}{1053–1061}.
\newblock
\showISBNx{9781450349147}
\urldef\tempurl%
\url{https://doi.org/10.1145/3041021.3055129}
\showDOI{\tempurl}


\bibitem[Macdiarmid et~al\mbox{.}(2016)]%
        {MACDIARMID2016EatingLikeTheresNoTomorrow}
\bibfield{author}{\bibinfo{person}{Jennie~I. Macdiarmid},
  \bibinfo{person}{Flora Douglas}, {and} \bibinfo{person}{Jonina Campbell}.}
  \bibinfo{year}{2016}\natexlab{}.
\newblock \showarticletitle{Eating like there's no tomorrow: Public awareness
  of the environmental impact of food and reluctance to eat less meat as part
  of a sustainable diet}.
\newblock \bibinfo{journal}{\emph{Appetite}}  \bibinfo{volume}{96}
  (\bibinfo{year}{2016}), \bibinfo{pages}{487 -- 493}.
\newblock
\showISSN{0195-6663}
\urldef\tempurl%
\url{https://doi.org/10.1016/j.appet.2015.10.011}
\showDOI{\tempurl}


\bibitem[Miller and Cassady(2015)]%
        {MILLER2015207}
\bibfield{author}{\bibinfo{person}{Lisa M.~Soederberg Miller} {and}
  \bibinfo{person}{Diana~L. Cassady}.} \bibinfo{year}{2015}\natexlab{}.
\newblock \showarticletitle{The effects of nutrition knowledge on food label
  use. A review of the literature}.
\newblock \bibinfo{journal}{\emph{Appetite}}  \bibinfo{volume}{92}
  (\bibinfo{year}{2015}), \bibinfo{pages}{207--216}.
\newblock
\showISSN{0195-6663}
\urldef\tempurl%
\url{https://doi.org/10.1016/j.appet.2015.05.029}
\showDOI{\tempurl}


\bibitem[Naturwaren(2022)]%
        {BNN}
\bibfield{author}{\bibinfo{person}{Bundesverband~Naturkost Naturwaren}.}
  \bibinfo{year}{2022}\natexlab{}.
\newblock \bibinfo{booktitle}{\emph{Planet-Score: Echte
  Nachhaltigkeitskennzeichnung statt Greenwashing}}.
\newblock
\urldef\tempurl%
\url{https://n-bnn.de/sites/default/dateien/bnn/sites/default/dateien/220323_BNN-Positionspapier_Planet-Score.pdf}
\showURL{%
\tempurl}


\bibitem[Nguyen et~al\mbox{.}(2019)]%
        {Nguyen2019_GreenConsumptionClosingIntentionBehaviorGap}
\bibfield{author}{\bibinfo{person}{Hung~Vu Nguyen}, \bibinfo{person}{Cuong~Hung
  Nguyen}, {and} \bibinfo{person}{Thoa Thi~Bao Hoang}.}
  \bibinfo{year}{2019}\natexlab{}.
\newblock \showarticletitle{Green consumption: Closing the intention-behavior
  gap}.
\newblock \bibinfo{journal}{\emph{Sustainable Development}}
  \bibinfo{volume}{27}, \bibinfo{number}{1} (\bibinfo{year}{2019}),
  \bibinfo{pages}{118--129}.
\newblock
\urldef\tempurl%
\url{https://doi.org/10.1002/sd.1875}
\showDOI{\tempurl}
\showeprint{https://onlinelibrary.wiley.com/doi/pdf/10.1002/sd.1875}


\bibitem[of~Consumer Co-operatives(2020)]%
        {HISPACOOP}
\bibfield{author}{\bibinfo{person}{European~Community of Consumer
  Co-operatives}.} \bibinfo{year}{2020}\natexlab{}.
\newblock \bibinfo{booktitle}{\emph{HISPACOOP Publishes Survey Findings on
  Front-Of-Pack Nutritional Labelling}}.
\newblock
\urldef\tempurl%
\url{https://www.eurocoop.coop/uploads/news/Hispacoop\%20NutriScore/HISPACOOP\%20Nutri-Score\%20Article.pdf}
\showURL{%
\tempurl}


\bibitem[Panzone et~al\mbox{.}(2021)]%
        {PANZONE2021nufood}
\bibfield{author}{\bibinfo{person}{Luca~A. Panzone}, \bibinfo{person}{Alistair
  Ulph}, \bibinfo{person}{Daniel~John Zizzo}, \bibinfo{person}{Denis Hilton},
  {and} \bibinfo{person}{Adrian Clear}.} \bibinfo{year}{2021}\natexlab{}.
\newblock \showarticletitle{The impact of environmental recall and carbon
  taxation on the carbon footprint of supermarket shopping}.
\newblock \bibinfo{journal}{\emph{Journal of Environmental Economics and
  Management}}  \bibinfo{volume}{109} (\bibinfo{year}{2021}),
  \bibinfo{pages}{102137}.
\newblock
\showISSN{0095-0696}
\urldef\tempurl%
\url{https://doi.org/10.1016/j.jeem.2018.06.002}
\showDOI{\tempurl}


\bibitem[Planet-Score(2023)]%
        {PlanetScore}
\bibfield{author}{\bibinfo{person}{Planet-Score}.}
  \bibinfo{year}{2023}\natexlab{}.
\newblock \bibinfo{booktitle}{\emph{Planet-Score}}.
\newblock
\urldef\tempurl%
\url{https://www.planet-score.org/en/}
\showURL{%
\tempurl}


\bibitem[Poore and Nemecek(2018)]%
        {Poore2018ReducingFoodImpact}
\bibfield{author}{\bibinfo{person}{J. Poore} {and} \bibinfo{person}{T.
  Nemecek}.} \bibinfo{year}{2018}\natexlab{}.
\newblock \showarticletitle{Reducing food{\textquoteright}s environmental
  impacts through producers and consumers}.
\newblock \bibinfo{journal}{\emph{Science}} \bibinfo{volume}{360},
  \bibinfo{number}{6392} (\bibinfo{year}{2018}), \bibinfo{pages}{987--992}.
\newblock
\showISSN{0036-8075}
\urldef\tempurl%
\url{https://doi.org/10.1126/science.aaq0216}
\showDOI{\tempurl}
\showeprint{https://science.sciencemag.org/content/360/6392/987.full.pdf}


\bibitem[Richet(2020)]%
        {LeMondeGreenWashing2}
\bibfield{author}{\bibinfo{person}{Enola Richet}.}
  \bibinfo{year}{2020}\natexlab{}.
\newblock \bibinfo{booktitle}{\emph{Le gouvernement accusé de faire du label
  haute valeur environnementale un « cheval de Troie du “greenwashing”
  »}}.
\newblock
\urldef\tempurl%
\url{https://www.lemonde.fr/planete/article/2020/12/16/le-gouvernement-accuse-de-faire-du-label-haute-valeur-environnementale-un-cheval-de-troie-du-greenwashing_6063589_3244.html}
\showURL{%
\tempurl}


\bibitem[Sanguinetti et~al\mbox{.}(2018)]%
        {SANGUINETTI2018_DesignBehaviourFrameworkEffectivenessEcoFeedback}
\bibfield{author}{\bibinfo{person}{Angela Sanguinetti}, \bibinfo{person}{Kelsea
  Dombrovski}, {and} \bibinfo{person}{Suhaila Sikand}.}
  \bibinfo{year}{2018}\natexlab{}.
\newblock \showarticletitle{Information, timing, and display: A design-behavior
  framework for improving the effectiveness of eco-feedback}.
\newblock \bibinfo{journal}{\emph{Energy Research \& Social Science}}
  \bibinfo{volume}{39} (\bibinfo{year}{2018}), \bibinfo{pages}{55 -- 68}.
\newblock
\showISSN{2214-6296}
\urldef\tempurl%
\url{https://doi.org/10.1016/j.erss.2017.10.001}
\showDOI{\tempurl}


\bibitem[{Santé publique France}(2020)]%
        {nslogo}
\bibfield{author}{\bibinfo{person}{{Santé publique France}}.}
  \bibinfo{year}{2020}\natexlab{}.
\newblock \bibinfo{booktitle}{\emph{Nutri-Score Charte graphique}}.
\newblock
\urldef\tempurl%
\url{https://www.santepubliquefrance.fr/media/files/02-determinants-de-sante/nutrition-et-activite-physique/nutri-score/annexe2-charte-graphique}
\showURL{%
\tempurl}


\bibitem[Sarda et~al\mbox{.}(2020)]%
        {Barthelemy2020nutriscoreFrance}
\bibfield{author}{\bibinfo{person}{Barthélemy Sarda}, \bibinfo{person}{Chantal
  Julia}, \bibinfo{person}{Anne-Juliette Serry}, {and} \bibinfo{person}{Pauline
  Ducrot}.} \bibinfo{year}{2020}\natexlab{}.
\newblock \showarticletitle{Appropriation of the Front-of-Pack Nutrition Label
  Nutri-Score across the French Population: Evolution of Awareness, Support,
  and Purchasing Behaviors between 2018 and 2019}.
\newblock \bibinfo{journal}{\emph{Nutrients}} \bibinfo{volume}{12},
  \bibinfo{number}{9} (\bibinfo{year}{2020}).
\newblock
\showISSN{2072-6643}
\urldef\tempurl%
\url{https://doi.org/10.3390/nu12092887}
\showDOI{\tempurl}


\bibitem[Satija et~al\mbox{.}(2017)]%
        {Satija2017veganheartdisease}
\bibfield{author}{\bibinfo{person}{Ambika Satija}, \bibinfo{person}{Shilpa~N.
  Bhupathiraju}, \bibinfo{person}{Donna Spiegelman},
  \bibinfo{person}{Stephanie~E. Chiuve}, \bibinfo{person}{JoAnn~E. Manson},
  \bibinfo{person}{Walter Willett}, \bibinfo{person}{Kathryn~M. Rexrode},
  \bibinfo{person}{Eric~B. Rimm}, {and} \bibinfo{person}{Frank~B. Hu}.}
  \bibinfo{year}{2017}\natexlab{}.
\newblock \showarticletitle{Healthful and Unhealthful Plant-Based Diets and the
  Risk of Coronary Heart Disease in U.S. Adults}.
\newblock \bibinfo{journal}{\emph{Journal of the American College of
  Cardiology}} \bibinfo{volume}{70}, \bibinfo{number}{4}
  (\bibinfo{year}{2017}), \bibinfo{pages}{411--422}.
\newblock
\urldef\tempurl%
\url{https://doi.org/10.1016/j.jacc.2017.05.047}
\showDOI{\tempurl}
\showeprint{https://www.jacc.org/doi/pdf/10.1016/j.jacc.2017.05.047}


\bibitem[Sauv\'{e} et~al\mbox{.}(2020)]%
        {sauve2020econundrum}
\bibfield{author}{\bibinfo{person}{Kim Sauv\'{e}}, \bibinfo{person}{Saskia
  Bakker}, {and} \bibinfo{person}{Steven Houben}.}
  \bibinfo{year}{2020}\natexlab{}.
\newblock \showarticletitle{Econundrum: Visualizing the Climate Impact of
  Dietary Choice through a Shared Data Sculpture}. In
  \bibinfo{booktitle}{\emph{Proceedings of the 2020 ACM Designing Interactive
  Systems Conference}} (Eindhoven, Netherlands) \emph{(\bibinfo{series}{DIS
  '20})}. \bibinfo{publisher}{Association for Computing Machinery},
  \bibinfo{address}{New York, NY, USA}, \bibinfo{pages}{1287–1300}.
\newblock
\showISBNx{9781450369749}
\urldef\tempurl%
\url{https://doi.org/10.1145/3357236.3395509}
\showDOI{\tempurl}


\bibitem[Scarborough et~al\mbox{.}(2014)]%
        {scarborough2014dietary}
\bibfield{author}{\bibinfo{person}{Peter Scarborough}, \bibinfo{person}{Paul~N
  Appleby}, \bibinfo{person}{Anja Mizdrak}, \bibinfo{person}{Adam~DM Briggs},
  \bibinfo{person}{Ruth~C Travis}, \bibinfo{person}{Kathryn~E Bradbury}, {and}
  \bibinfo{person}{Timothy~J Key}.} \bibinfo{year}{2014}\natexlab{}.
\newblock \showarticletitle{Dietary greenhouse gas emissions of meat-eaters,
  fish-eaters, vegetarians and vegans in the UK}.
\newblock \bibinfo{journal}{\emph{Climatic change}} \bibinfo{volume}{125},
  \bibinfo{number}{2} (\bibinfo{year}{2014}), \bibinfo{pages}{179--192}.
\newblock


\bibitem[Shakeri and McCallum(2021)]%
        {shakeri2021envirofy}
\bibfield{author}{\bibinfo{person}{G\"{o}zel Shakeri} {and}
  \bibinfo{person}{Claire~H McCallum}.} \bibinfo{year}{2021}\natexlab{}.
\newblock \showarticletitle{Envirofy Your Shop: Development of a Real-Time Tool
  to Support Eco-Friendly Food Purchases Online}. In
  \bibinfo{booktitle}{\emph{Extended Abstracts of the 2021 CHI Conference on
  Human Factors in Computing Systems}} (Yokohama, Japan)
  \emph{(\bibinfo{series}{CHI EA '21})}. \bibinfo{publisher}{Association for
  Computing Machinery}, \bibinfo{address}{New York, NY, USA}, Article
  \bibinfo{articleno}{390}, \bibinfo{numpages}{10}~pages.
\newblock
\showISBNx{9781450380959}
\urldef\tempurl%
\url{https://doi.org/10.1145/3411763.3451713}
\showDOI{\tempurl}


\bibitem[Simeone et~al\mbox{.}(2016)]%
        {Simeone2016labelcomplexity}
\bibfield{author}{\bibinfo{person}{Mariarosaria Simeone},
  \bibinfo{person}{Debora Scarpato}, {and} \bibinfo{person}{Nicola Marinelli}.}
  \bibinfo{year}{2016}\natexlab{}.
\newblock \showarticletitle{Factors Affecting Food Label Complexity: Does the
  New EU Regulation Satisfy Consumer Issues? An Exploratory Study}.
\newblock \bibinfo{journal}{\emph{Journal of Food Products Marketing}}
  \bibinfo{volume}{22}, \bibinfo{number}{5} (\bibinfo{year}{2016}),
  \bibinfo{pages}{571--583}.
\newblock
\urldef\tempurl%
\url{https://doi.org/10.1080/10454446.2015.1121422}
\showDOI{\tempurl}
\showeprint{https://doi.org/10.1080/10454446.2015.1121422}


\bibitem[Southey(2021)]%
        {FOPfrance}
\bibfield{author}{\bibinfo{person}{Flora Southey}.}
  \bibinfo{year}{2021}\natexlab{}.
\newblock \bibinfo{booktitle}{\emph{First Nutri-Score for nutrition, now
  Eco-Score for the environment: New FOP lands in France}}.
\newblock
\urldef\tempurl%
\url{https://www.foodnavigator.com/Article/2021/01/12/Eco-Score-New-FOP-label-measures-the-environmental-impact-of-food}
\showURL{%
\tempurl}


\bibitem[Starke et~al\mbox{.}(2017)]%
  {Starke2017_EffectiveUserInterfaceDesignsIncreaseEnergyEfficientBehaviour}
\bibfield{author}{\bibinfo{person}{Alain Starke}, \bibinfo{person}{Martijn
  Willemsen}, {and} \bibinfo{person}{Chris Snijders}.}
  \bibinfo{year}{2017}\natexlab{}.
\newblock \showarticletitle{Effective User Interface Designs to Increase
  Energy-Efficient Behavior in a Rasch-Based Energy Recommender System}. In
  \bibinfo{booktitle}{\emph{Proceedings of the Eleventh ACM Conference on
  Recommender Systems}} (Como, Italy) \emph{(\bibinfo{series}{RecSys '17})}.
  \bibinfo{publisher}{Association for Computing Machinery},
  \bibinfo{address}{New York, NY, USA}, \bibinfo{pages}{65–73}.
\newblock
\showISBNx{9781450346528}
\urldef\tempurl%
\url{https://doi.org/10.1145/3109859.3109902}
\showDOI{\tempurl}


\bibitem[Terzimehi\'{c} et~al\mbox{.}(2017)]%
        {Terzimehi2017_ChallengesJustInTimeAdaptiveFoodChoiceInterventions}
\bibfield{author}{\bibinfo{person}{Naundefineda Terzimehi\'{c}},
  \bibinfo{person}{Christina Schneegass}, {and} \bibinfo{person}{Heinrich
  Hu\ss{}mann}.} \bibinfo{year}{2017}\natexlab{}.
\newblock \showarticletitle{Exploring Challenges in Automated Just-In-Time
  Adaptive Food Choice Interventions}. In \bibinfo{booktitle}{\emph{Proceedings
  of the 2nd International Workshop on Multimedia for Personal Health and
  Health Care}} (Mountain View, California, USA)
  \emph{(\bibinfo{series}{MMHealth '17})}. \bibinfo{publisher}{Association for
  Computing Machinery}, \bibinfo{address}{New York, NY, USA},
  \bibinfo{pages}{81–84}.
\newblock
\showISBNx{9781450355049}
\urldef\tempurl%
\url{https://doi.org/10.1145/3132635.3132648}
\showDOI{\tempurl}


\bibitem[Test-Achats(2020)]%
        {TestAchats}
\bibfield{author}{\bibinfo{person}{Test-Achats}.}
  \bibinfo{year}{2020}\natexlab{}.
\newblock \bibinfo{booktitle}{\emph{Sondage de Test Achats une majorité est
  favorable à un Nutri-Score obligatoire}}.
\newblock
\urldef\tempurl%
\url{https://www.test-achats.be/sante/alimentation-et-nutrition/alimentation-saine/presse/2020/sondage-de-test-achats-une-majorite-est-favorable-a-un-nutri-score-obligatoire}
\showURL{%
\tempurl}


\bibitem[Torma and Thøgersen(2021)]%
        {TORMA2021metalabelling}
\bibfield{author}{\bibinfo{person}{Gabriele Torma} {and} \bibinfo{person}{John
  Thøgersen}.} \bibinfo{year}{2021}\natexlab{}.
\newblock \showarticletitle{A systematic literature review on meta
  sustainability labeling – What do we (not) know?}
\newblock \bibinfo{journal}{\emph{Journal of Cleaner Production}}
  \bibinfo{volume}{293} (\bibinfo{year}{2021}), \bibinfo{pages}{126194}.
\newblock
\showISSN{0959-6526}
\urldef\tempurl%
\url{https://doi.org/10.1016/j.jclepro.2021.126194}
\showDOI{\tempurl}


\bibitem[Vermeir and Verbeke(2006)]%
  {Vermeir2006_SustainableFoodConsumptionExploringConsumerAttitudeBehavioralIntentionGap}
\bibfield{author}{\bibinfo{person}{Iris Vermeir} {and} \bibinfo{person}{Wim
  Verbeke}.} \bibinfo{year}{2006}\natexlab{}.
\newblock \showarticletitle{Sustainable food consumption: Exploring the
  consumer “attitude--behavioral intention” gap}.
\newblock \bibinfo{journal}{\emph{Journal of Agricultural and Environmental
  ethics}} \bibinfo{volume}{19}, \bibinfo{number}{2} (\bibinfo{year}{2006}),
  \bibinfo{pages}{169--194}.
\newblock


\bibitem[Yalva\c{c} et~al\mbox{.}(2014)]%
        {yalvac2014socialrecipes}
\bibfield{author}{\bibinfo{person}{Fulya Yalva\c{c}}, \bibinfo{person}{Veranika
  Lim}, \bibinfo{person}{Jun Hu}, \bibinfo{person}{Mathias Funk}, {and}
  \bibinfo{person}{Matthias Rauterberg}.} \bibinfo{year}{2014}\natexlab{}.
\newblock \showarticletitle{Social Recipe Recommendation to Reduce Food Waste}.
  In \bibinfo{booktitle}{\emph{CHI '14 Extended Abstracts on Human Factors in
  Computing Systems}} (Toronto, Ontario, Canada) \emph{(\bibinfo{series}{CHI EA
  '14})}. \bibinfo{publisher}{Association for Computing Machinery},
  \bibinfo{address}{New York, NY, USA}, \bibinfo{pages}{2431–2436}.
\newblock
\showISBNx{9781450324748}
\urldef\tempurl%
\url{https://doi.org/10.1145/2559206.2581226}
\showDOI{\tempurl}


\bibitem[Yokessa and Marette(2019)]%
        {yokessa:hal-02628579}
\bibfield{author}{\bibinfo{person}{Ma{\"i}mouna Yokessa} {and}
  \bibinfo{person}{Stephan~S. Marette}.} \bibinfo{year}{2019}\natexlab{}.
\newblock \showarticletitle{{A Review of Eco-labels and their Economic
  Impact}}.
\newblock \bibinfo{journal}{\emph{{International Review of Environmental and
  Resource Economics}}} \bibinfo{volume}{13}, \bibinfo{number}{1-2}
  (\bibinfo{year}{2019}), \bibinfo{pages}{119--163}.
\newblock
\urldef\tempurl%
\url{https://doi.org/10.1561/101.00000107}
\showDOI{\tempurl}


\bibitem[Zapico et~al\mbox{.}(2016)]%
        {zapico2016ecopanel}
\bibfield{author}{\bibinfo{person}{Jorge~Luis Zapico}, \bibinfo{person}{Cecilia
  Katzeff}, \bibinfo{person}{Ulrica Bohn\'{e}}, {and} \bibinfo{person}{Rebecka
  Milestad}.} \bibinfo{year}{2016}\natexlab{}.
\newblock \showarticletitle{Eco-Feedback Visualization for Closing the Gap of
  Organic Food Consumption}. In \bibinfo{booktitle}{\emph{Proceedings of the
  9th Nordic Conference on Human-Computer Interaction}} (Gothenburg, Sweden)
  \emph{(\bibinfo{series}{NordiCHI '16})}. \bibinfo{publisher}{Association for
  Computing Machinery}, \bibinfo{address}{New York, NY, USA}, Article
  \bibinfo{articleno}{75}, \bibinfo{numpages}{9}~pages.
\newblock
\showISBNx{9781450347631}
\urldef\tempurl%
\url{https://doi.org/10.1145/2971485.2971507}
\showDOI{\tempurl}


\end{thebibliography}

\appendix
\newpage
\section{Nutri-Score}\label{app:a}
\begin{figure}[h!]
\centering
\includegraphics[width=\textwidth]{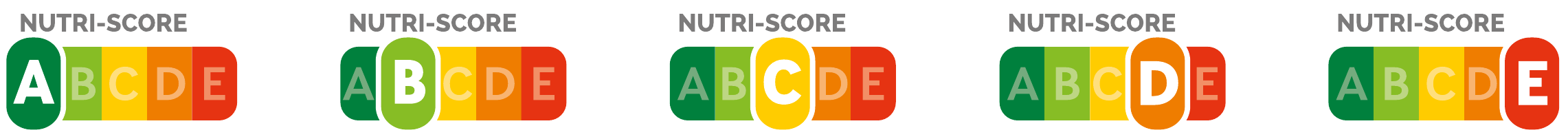}
\caption{Nutri-Score labels \cite{off}}
\label{fig:nutri-score} 
\Description{Figure 5 shows all five variants of the Nutri-Score label with the ratings `A' to `E'.}
\end{figure}

\section{Eco-Score}\label{app:b}
\begin{figure}[h!]
\centering
\includegraphics[width=\textwidth]{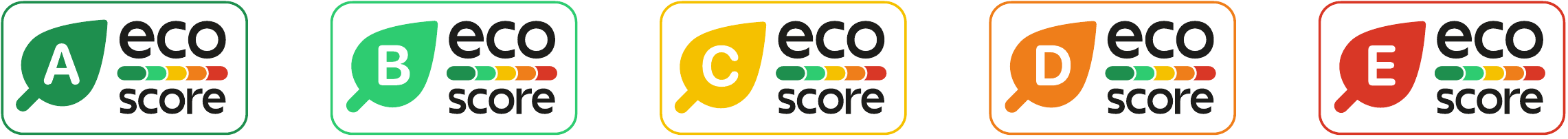}
\caption{Eco-Score labels \cite{off}}
\label{fig:eco-score}
\Description{Figure 6 shows all five variants of the Eco-Score label with the ratings `A' to `E'.}
\end{figure}

\section{Statistics of gathering requirements process}\label{app:c}
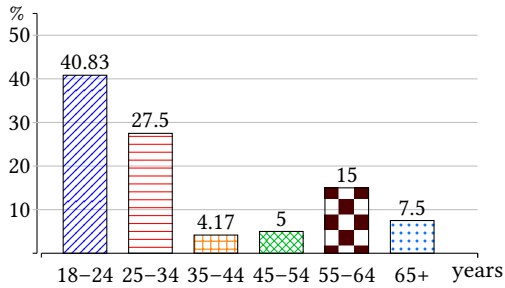
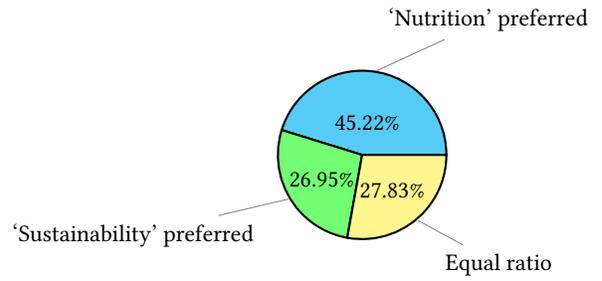
\begin{figure}[h!]%
     \centering
     \subfloat[{\centering Age groups of the participants\label{fig:age-stat}}]{\begin{tikzpicture}[scale=0.58]
   \draw (0cm,0cm) -- (10cm,0cm);  %Abzisse
   \draw (0cm,0cm) -- (0cm,-0.1cm);  %linkes Ende der Abzisse
   \draw (10cm,0cm) -- (10cm,-0.1cm);  %rechtes Ende der Abzisse
   \draw (10cm,0cm) -- (10cm,-0.1cm) node [below] {years};
  
   \draw (-0.1cm,0cm) -- (-0.1cm,5.5cm);  %Ordinate
   \draw (-0.1cm,0cm) -- (-0.2cm,0cm);  %unteres Ende der Ordinate
   \draw (-0.1cm,5.5cm) -- (-0.2cm,5.5cm) node [left] {\%};  %oberes Ende der Ordinate

   \draw[gray!50, text=black] (-0.2cm,1cm) -- (10cm,1cm) node at (-0.5cm,1cm) {10};
   \draw[gray!50, text=black] (-0.2cm,2cm) -- (10cm,2cm) node at (-0.5cm,2cm) {20};
   \draw[gray!50, text=black] (-0.2cm,3cm) -- (10cm,3cm) node at (-0.5cm,3cm) {30};
   \draw[gray!50, text=black] (-0.2cm,4cm) -- (10cm,4cm) node at (-0.5cm,4cm) {40};
   \draw[gray!50, text=black] (-0.2cm,5cm) -- (10cm,5cm) node at (-0.5cm,5cm) {50};

   %\node at (5.5cm,5.5cm) {Age groups of the participants};  %Überschrift
                           
   \draw[pattern=north east lines, pattern color=blue] (0.5cm,0cm) rectangle (1.5cm,40.83mm)
       node at (1cm,43.83mm) {40.83};   
   \node at (1cm, -0.5cm) {18--24};
   \draw[pattern=horizontal lines, pattern color=red] (2cm,0cm) rectangle (3cm,27.5mm)
       node at (2.5cm,30.5mm) {27.5};   
   \node at (2.5cm, -0.5cm) {25--34};
   \draw[pattern=grid, pattern color=orange] (3.5cm,0cm) rectangle (4.5cm,4.17mm)
       node at (4cm,7.17mm) {4.17};   
   \node at (4cm, -0.5cm) {35--44};
   \draw[pattern=crosshatch, pattern color=PineGreen] (5cm,0cm) rectangle (6cm,5mm)
       node at (5.5cm,8mm) {5};   
   \node at (5.5cm, -0.5cm) {45--54};
   \draw[pattern=checkerboard, pattern color=Sepia] (6.5cm,0cm) rectangle (7.5cm,15mm)
       node at (7cm,18mm) {15};   
   \node at (7cm, -0.5cm) {55--64};
   \draw[pattern=dots, pattern color=NavyBlue] (8cm,0cm) rectangle (9cm,7.5mm)
       node at (8.5cm,10.5mm) {7.5};   
   \node at (8.5cm, -0.5cm) {65+};
   %\draw[pattern=checkerboard, pattern color=Sepia] (9.5cm,0cm) rectangle (10.5cm,1.47mm)
       %node at (10cm,4.47mm) {1.47};   
   %\node at (10cm, -0.5cm) {N/A};
\end{tikzpicture}}%
\quad
\subfloat[{\centering Desired prioritisations of nutrition and sustainability information in the final label \label{fig:prio-stat} \cite[adjusted presentation]{thesis}}]{\begin{tikzpicture}[scale=0.35]
\pie[color={cyan!55, green!55, yellow!55},radius=3.2, text=pin]{45.22/`Nutrition' preferred, 26.95/`Sustainability' preferred, 27.83/Equal ratio}
\end{tikzpicture}}%
\caption{Age demographics and prioritisation preferences based on an online survey with 120 Germans}
\label{fig:stats}%
\Description{Figure 7 labelled `(a)' contains a bar chart showing the percentage share of the age groups of the participants of our online survey. The shares are: 1. 40.83\% participants aged between 18 to 24 years. 2. 27.5\% participants aged between 25 to 34 years. 3. 4.17\% participants aged between 35 to 44 years. 4. 5\% participants aged between 45 to 54 years. 5. 15\% participants aged between 55 to 64 years. 6. 7.5\% participants aged at least 65 years. Figure 7 labelled `(b)' contains a pie chart showing the percentage share of the participants' prioritisations regarding the parameters `Nutrition' and `Sustainability'. The participants stated that 1. 45.22\% prioritise Nutrition over Sustainability, 2. 26.95\% prioritise Sustainability over Nutrition, 3. 27.83\% stated an equal ratio.}
\end{figure}

\newpage %temporary

\section{Scale-Score}\label{app:d}
\begin{figure}[h!]
\centering
\includegraphics[width=\textwidth]{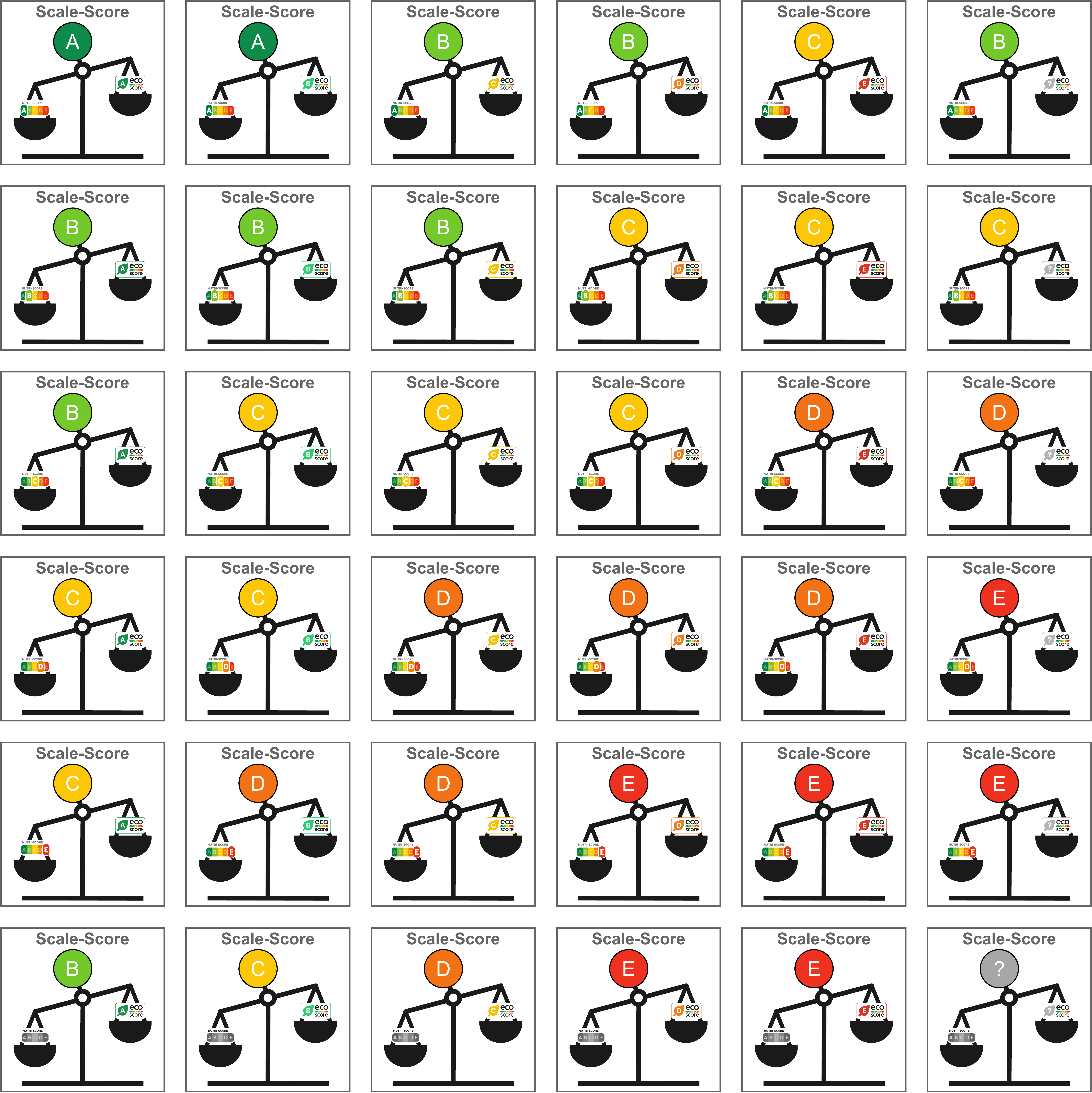}
\caption{Combining label: Scale-Score (all variants) \cite{thesis}}
\label{fig:scale-variants}
\Description{Figure 8 shows all 36 variants of the Scale-Score label, based on a product's Nutri- and Eco-Score.}
\end{figure}

\newpage %temporary

\section{Software architecture}\label{app:e}
\begin{figure}[h!]
\centering
\includegraphics[width=0.9\textwidth]{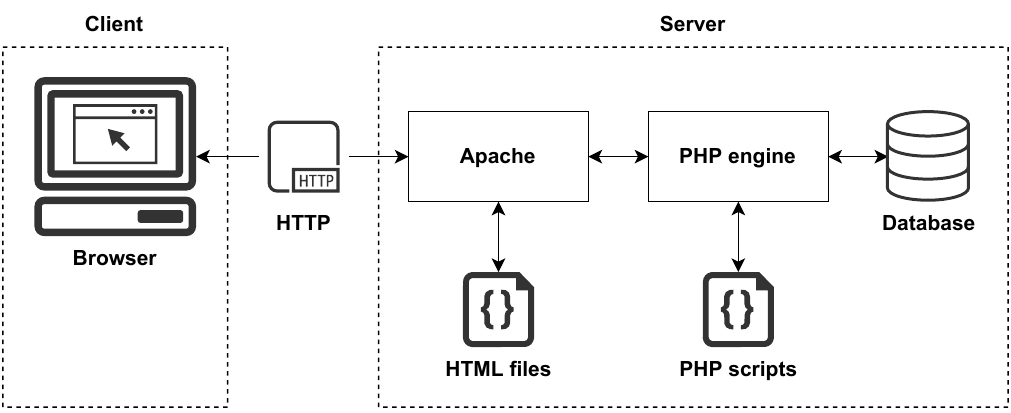}
\caption{Used software architecture \cite[adjusted presentation]{xampp,thesis}}
\label{fig:xampp}
\Description{Figure 9 shows the software architecture of the developed online store. It shows the communication via HTTP between the two components Client and Server. On the client side, only the browser is shown. On the server side, the communication between the Apache Webserver, managing the implemented HTML files and the PHP engine is shown. PHP scripts are interpreted by this engine, managing the database access in our online store.}
\end{figure}

\section{Study set-up}\label{app:f}
\begin{figure}[h!]
\centering
\includegraphics[width=0.8\textwidth]{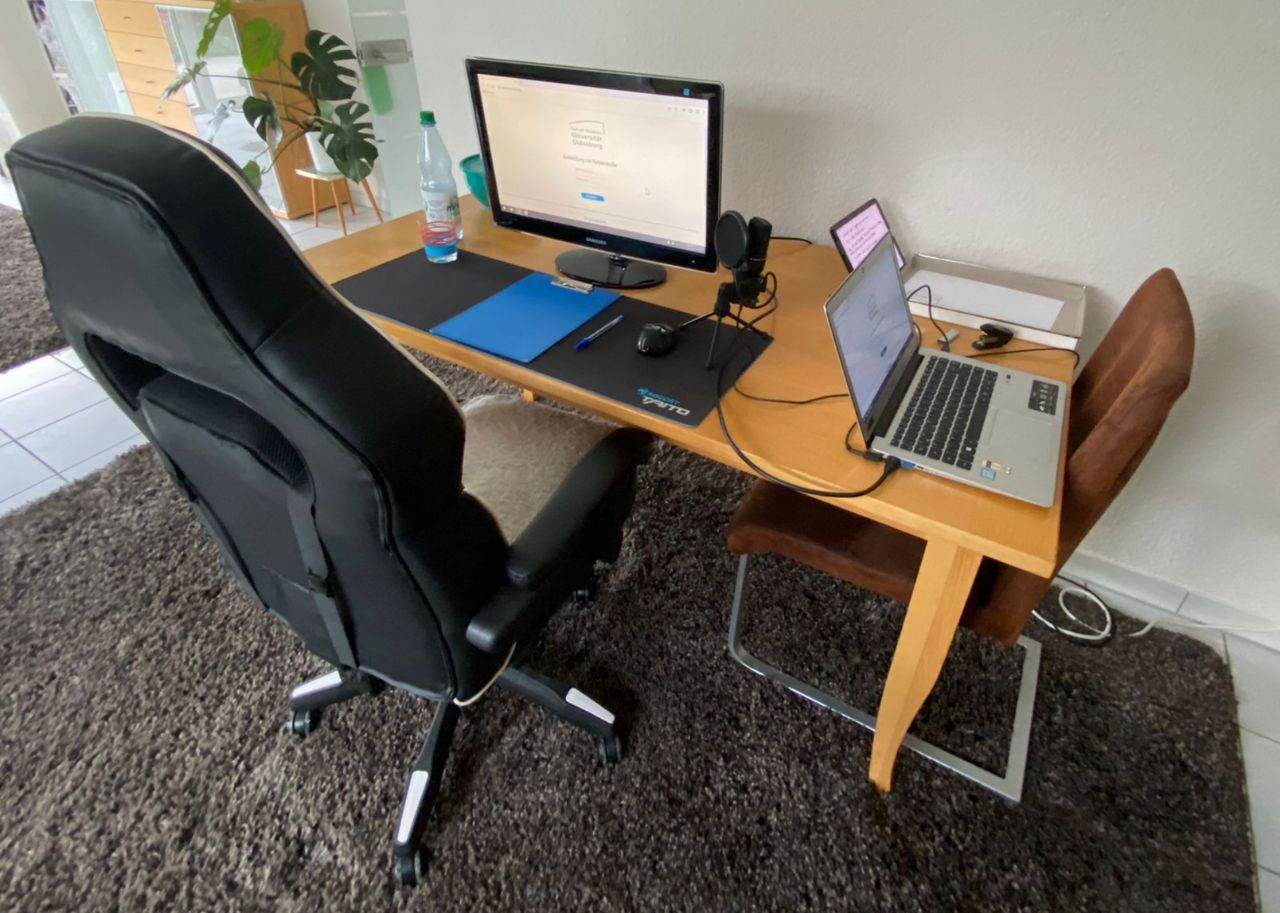}
\caption{Laboratory environment \cite{thesis}. Due to COVID-19 restrictions, a laboratory environment was recreated in the first author's living room. This study was conducted in accordance with COVID-19 regulations valid at the time. A height adjustable and padded chair was provided for the comfort of the participants.}
\label{fig:aufbau}
\Description{Figure 10 shows our laboratory environment. Participants were seated in front of a desk with a personal computer in front of a white wall. Pictures or posters were removed from the wall so as not to distract the participants.}
\end{figure}

\end{document}